A new, optimized Doppler optical probe for phase detection, bubble velocity and size measurements: investigation of a bubble column operated in the heterogeneous regime.

Anthony Lefebvre[a]*, Yann Mezui[b], Martin Obligado[b], Stéphane Gluck[a], Alain Cartellier[b]

[a] A2 Photonic Sensors, 38016 Grenoble, France
[b] Univ. Grenoble Alpes, CNRS, Grenoble INP**, *LEGI*, 38000 Grenoble, France

* Corresponding author: Anthony Lefebvre, alefebvre@a2photonicsensors.com
** Institute of Engineering Univ. Grenoble Alpes.

**Highlights**
- A new optical sensor for bubble velocity and size measurements is proposed.
- The sensor gives the bubble translation velocity projected along the fiber axis.
- Reliable data are collected in a bubble column up to 30% global gas holdup.
- A self-similar bubble velocity profile is established in the heterogeneous regime.
- Bubble turbulent vertical intensity reached 55% in the heterogeneous regime.

**ABSTRACT**
A new measuring technique dedicated to bubble velocity and size measurements in complex bubbly flows such as those occurring in bubble columns is proposed. This sensor combines the phase detection capability of a conical optical fiber, with velocity measurements from the Doppler signal induced by an interface approaching the extremity of a single-mode fiber. The analysis of the probe functioning and of its response in controlled situations, have shown that the Doppler probe provides the translation velocity of bubbles projected along the probe axis. A reliable signal processing routine has been developed that exploits the Doppler signal arising at the gas-to-liquid transition: the resulting uncertainty on velocity is at most 14%. Such a Doppler probe provides statistics on velocity and on size of gas inclusions, as well as local variables including void fraction, gas volumetric flux, number density and its flux. That sensor has been successfully exploited in an air-tap water bubble column 0.4m in diameter for global gas hold-up from 2.5 to 30%. In the heterogeneous regime, the transverse profiles of the mean bubble velocity scaled by the value on the axis happen to be self-similar in the quasi fully developed region of the column. A fit is proposed for these profiles. In addition, on the axis, the standard deviation of bubble velocity scaled by the mean velocity increases with Vsg in the homogeneous regime, and it remains stable, close to 0.55, in the heterogeneous regime.

**1. Introduction**

Bubble column reactors, where gas is injected at the bottom of an initially stagnant liquid, are widely used in chemical engineering (e.g. hydrogenation such as in Fischer-Tropsch synthetic fuels production), for bio-chemical transformations (e.g. in aerobic bioreactors, for algae production…), in waste management (e.g. wet oxidation) or for flotation units (e.g. deinking of paper, extraction of rare metals…). These reactors are usually operated in the so-called heterogeneous regime for which the gas concentration typically ranges from 15% to 40%. The resulting hydrodynamics is complex. A mean large-scale motion takes place in the column consisting of an upward motion of the liquid with a high gas concentration in the center of the column, compensated by a downward motion of the liquid with a lower gas concentration near walls. That bubble driven fluid circulation at the reactor scale has been identified in the sixties, notably by De Nevers, 1968 and by Pavlov and Pozin quoted in Hills, 1974. In addition, De Nevers, 1968 pointed out that "these circulations are unstable and change size, shape and orientation chaotically" and that they are the "principal mode of vertical bubble transport". These unsteady motions arise not only at the reactor scale but also at smaller length and time scales due to the formation of eddies and of "spiral-like structures" (e.g. Chen et al., 1994; Ruzicka et al., 2001 and references therein). Owing to such complexity, and in spite of continuous research efforts since the 60's, the hydrodynamics of bubble columns still remains poorly understood. This is manifest from the lack of consensus on scale-up rules for bubble columns as demonstrated by the number and the variety of correlations that have been proposed so far (e.g. Deckwer, 1992;

Joshi et al., 1998; Kantarci et al., 2005; Rollbusch et al., 2015; Kikukawa, 2017; Besagni et al., 2018 to quote a few).

Similarly, numerical simulations of bubble columns have not yet reached a predictive status since *ad-hoc* adjustments are still required notably when the column diameter is varied (e.g. Ekambara et al., 2005). Recently, the situation regarding scale-up and simulation performances has started to improve thanks to well controlled experiments performed over an extended range of column diameter (from 0.1 up to 3m) and of gas superficial velocities, and also with the help of a new measuring technique that provides the Sauter mean diameter of bubbles evaluated along the horizontal direction (Maximiano Raimundo et al., 2016). Various new features have been identified from these experiments. The presence of powerful unsteady motions has been confirmed at all column scales: these motions lead to turbulent intensities about 25-30% on the column axis that grow up to 70% at a radial distance equal to 0.7 column radius (Maximiano Raimundo et al., 2019). These large fluctuations happen to be a key characteristic of the heterogeneous regime compared with the homogeneous one, and we tentatively attributed their origin to the convective instabilities driven by the strong concentration gradients present in these flows. Indeed, regions with void fraction up to ten times the mean hold-up (denoted as clusters) and down to 0.1 times the mean hold-up (denoted as voids) have been recently detected in bubble columns operated in the heterogeneous regime (Maximiano Raimundo, 2015; Maximiano Raimundo et al., 2019). The presence of such concentration gradients has many consequences. Since these concentration gradients induce buoyancy fluctuations, they are prone to favor localized shear and hence to drive turbulence production in bubble columns. They are also expected to enhance the apparent relative velocity between phases as bubbles are preferentially encountered in clusters while the liquid phase is mostly present in voids. Such an enhancement has indeed been detected in experiments (see for example Maximiano Raimundo, 2015; Maximiano Raimundo et al., 2019), and it is now accounted for in simulations (e.g. Maximiano Raimundo, 2015; Gemello et al., 2018) notably by introducing a swarm factor to correct the drag force such as the one proposed by Simonnet et al., 2007. McLure et al., 2017 have recently proposed alternate swarm factors that have the peculiarity to depend both on void fraction and on the mean bubble size. Accounting for such a collective dynamics for bubbles allow two-fluid models to better represent interfacial momentum exchanges. In particular, when including an adapted swarm factor, gas hold-up and liquid velocity on the column axis agree within ±15% with experiments (Maximiano Raimundo, 2015; Gemello et al., 2018) and over a significant range of flow conditions (namely for column diameter from 0.12 to 3m and for superficial velocity from 3 to 35cm/s). Furthermore, that apparent relative velocity has been shown to exceed the terminal velocity of individual bubbles by a factor up to 3 (Maximiano Raimundo et al., 2019, Maximiano Raimundo, 2015), indicating that the bubble terminal velocity does not control the velocity scale. Instead, the quantity $(gD)^{1/2}$ has been shown to be a natural velocity scale for the liquid phase in bubble columns and as such, it should appear in the scale-up rules (Maximiano Raimundo et al., 2019). Such a scale is reminiscent of turbulent flows driven by convection, a feature that emphasizes further the similarity between confined turbulent convection and bubble columns in the heterogeneous regime (Maximiano Raimundo et al., 2019).

So far, our understanding of bubble column dynamics relies on local measurements of void fraction, bubble size and of velocity and Reynolds stress in the liquid phase. Information regarding gas velocity statistics in the heterogeneous regime is extremely rare. In particular, it is not established whether or not the scale $(gD)^{1/2}$ also applies to the gas phase. Fluctuations in the gas velocity that contribute to bubble induced turbulence are not documented, and, owing to the discussion above, the connection between meso-scale structures and the gas phase relative velocity needs to be clarified. In that perspective, and as done for turbulent laden flows (Aliseda et al., 2002; Sumbekova et al., 2016; Huck et al., 2018), measurements of bubble velocities distributions conditioned by the local void fraction are highly desirable to connect the collective dynamics of bubbles to the characteristics of cluster and void regions. Clearly, to improve our understanding of momentum exchange mechanisms occurring in dense bubble columns as well as their representation in two-fluids models, there is a crucial need to gather reliable statistics on bubble velocities. The present lack of information arises from limitations of available measuring technique. As imaging or laser based techniques are not suited to investigate dense bubbly flows, previous attempts have exploited intrusive sensors. Single as well as double optical probes happen to provide erroneous velocity distributions when used in the heterogeneous regime because of bubbles incoming from the sides of the probe or from its rear. Indeed, inclined trajectories induce small de-wetting times on mono-fiber probes (Maximiano Raimundo et al., 2016) and small time of flight between two tips of double (Chaumat et al., 2007) or of multiple probes (Xue et al., 2003): in all cases, large, unphysical interface velocities are recorded. The above techniques are therefore too sensitive to the bubble trajectory. Four-tips probes have been designed to better discriminate between trajectories. Indeed, four-tips probes provide three times-of-flight corresponding to interface displacements from the central tip to each of the three other tips, and comparing these times-of-flight allows to recognize and to remove bubble trajectories at an angle with the probe axial direction. Four-tips sensors are subject to two main limitations. First, they are efficient on bubbles larger than about twice the lateral distance between probe tips. Efforts have been devoted to diminish that lateral distance down to about 0.5-0.6mm (see Xue et al., 2003; Mudde and Saito, 2001; Saito and Mudde, 2001; Guet et al., 2003) and even 0.25mm (Sakamoto and Saito, 2012). The minimum vertical dimension of bubbles properly detected so far with four-tips probes when used in elementary flows (e.g. isolated bubble, train of bubbles…) is about 1.4 to 1.6mm. In more realistic two-phase flows, when a bubble size below these limits is detected, it is usually discarded from the statistics (Mudde and Saito, 2001). For actual two-phase flows, a second limitation arises in connection with information processing. Ideally, identical times-of-flight correspond to bubbles centered on the probe and whose velocity is directed along the probe axis. Hence, enforcing a strong criterion on the similarity between the time-of-flight drastically diminishes the fraction of bubbles detected and accounted for in the measurement (Guet et al., 2003): according to Xue et al., 2003, that decrease can be as large as 99%. In practice, the similarity criterion is relaxed and its value is selected as a balance between the rise of the data collection rate and the increasing uncertainty. However, relaxing that criterion leads to some - hard to evaluate - statistical bias as the bubble population is not uniformly scrutinized. Various improvements have been brought on the signal processing to try to circumvent these drawbacks. In particular, to properly demodulate the collected information, one has to account for bubble curvature and shape possibly including oscillations, for its orientation, position and trajectory with respect to the probe. Consequently, all these variables also contribute to uncertainty sources, and they come in addition to uncertainties arising from bubble-probe interactions. The later include bubble deceleration, interface deformation and trajectory distortions notably because of crawling. Harteveld, 2005 carefully investigated uncertainty sources for four–tip probes and concluded that the mean bubble velocity is usually well captured, but the reliability of gas velocity fluctuations measurement is far from being ascertained. Owing to the discussion above, it is not surprising that four-tips sensors have been mostly used in quasi one-dimensional flows where the uncertainty due to bubble trajectory remains limited. To our knowledge, the only exception is the investigation of churn-turbulent (i.e. heterogeneous) conditions in a 0.16m I.D. bubble column by Xue et al., 2008. These authors considered a large range of superficial velocities (from 2 to 60cm/s) leading to local void fractions up to 80%. At a given position in the column, a four-tip sensor (with a 0.6mm lateral distance between tips) was oriented vertically upward to gather positive bubble velocities, and then vertically downward to collect negative bubble velocities: the reconstructed gas velocity distributions exhibited two peaks corresponding to upward and downward motions. Chord length distributions as well as chord-velocity correlations were also collected. As the distributions included a zero velocity, Xue et al., 2008 argue that a significant bias may occur in the vicinity of zero velocity as the gas dwell time increases and may become larger than the time scale associated with the change in the flow direction. Also, part of the uncertainty in velocity measurements was attributed to the oscillations of the interface as large bubbles, up to a few centimeters in size, were present in the flow. Aside these two comments, the uncertainty was not quantified, but the authors show that the original results they obtained exhibit consistent trends when varying the gas superficial velocity. Oddly, Xue et al., 2008 do not provide gas velocity fluctuations or local gas fluxes.

In this context, our objective was to develop a sensor able to provide reliable information on bubble velocities statistics and on gas flux in the challenging conditions encountered in a bubble column operated in conditions relevant for industry. The main issue was to ensure a strong sensitivity of the sensor response to bubble trajectory in order to discriminate meaningful signatures even for distorted gas inclusions experiencing 3D unsteady motions. To do so, we started from conical optical probes whose reliability in terms of interface detection and void fraction determination has been proved accurate even in the difficult flow conditions encountered in the heterogeneous regime (Maximiano Raimundo et al., 2016), and we developed a new sensor that systematically exploits the Doppler signals collected with a mono-mode optical fiber.

**2. Principle of operation, new sensors and their performances on controlled interface**

To illustrate the principle of velocity measurements, let us consider an optical fiber with one extremity immersed in the two-phase flow. Coherent light (frequency $f_0$, wavelength in vacuum $\lambda_0$) is injected in the fiber. The fraction of incoming light reflected at the immersed tip depends on the refractive index of the media that surrounds it: this is the principle exploited for phase detection when using optical probes (e.g. Cartellier and Achard, 1991; Cartellier, 2001). To access the interface velocity V, we considered the Doppler signal formed by combining the wave reflected at the fiber extremity that has the same frequency as the light source, and the wave reflected by the approaching interface in the outer medium, that re-enters the fiber and whose frequency has been Doppler shifted. Let us start with the idealized situation of a plane wave that encounters a reflector approaching at a velocity V along the direction of propagation of the light beam (Fig.1 with an angle $\alpha$=0). In a fixed frame of reference, the frequency of the reflected wave equals $f_0 - 2V/\lambda$ where $\lambda = \lambda_0/n_{ext}$ is the wavelength in the medium of propagation those refractive index is $n_{ext}$. The beating of the two waves at frequencies $f_0$ and $f_0 - 2V/\lambda$ provides at the receiver a signal modulated at the Doppler frequency $f_D = 2V/(\lambda_0/n_{ext})$. The reflector velocity can thus be inferred by measuring the Doppler frequency $f_D$. Note that the relevant refractive index $n_{ext}$ is that of the medium into which the light propagates. For example, for the front interface of a bubble approaching the sensor, $n_{ext}$ must be taken as the refractive index of the liquid. For a probe exiting a gas bubble or for a drop approaching the probe in a gas, $n_{ext}$ is the refractive index of the gas. When the velocity of the approaching reflector is at an angle $\alpha$ with the direction of the light propagation, the velocity component to be accounted for is the projection of the interface displacement velocity (by definition directed along the normal **n** to the interface) on the direction **k** of the incoming wave, i.e. $|V \mathbf{n}.\mathbf{k}| = V \cos(\alpha)$. Hence, the Doppler frequency expresses as:

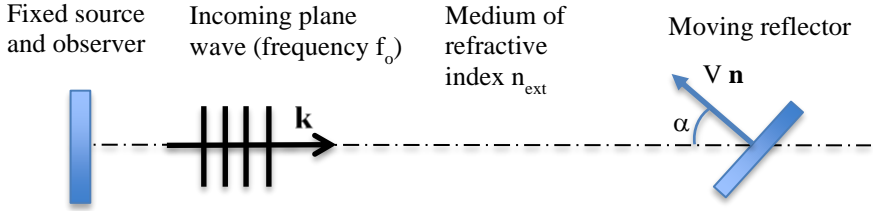

$$f_D = 2\frac{V|\mathbf{n}\cdot\mathbf{k}|}{\lambda_0/n_{ext}} \qquad (1)$$

Fixed source and observer | Incoming plane wave (frequency $f_o$) | Medium of refractive index $n_{ext}$ | Moving reflector $V\mathbf{n}$, $\alpha$

Fig.1: Idealized situation of a planar wave directed along the vector **k** interacting with a reflector moving at a speed V along its normal **n**.

Such a principle has been or exploited a few times in multiphase flows. Using cleaved single mode as well as multimode optical fibers, Podkorytov et al., 1989 detected Doppler signals due to approaching interfaces during their investigation of boiling in cryogenic fluids. Doppler signals collected with a multimode optical fiber (50µm core diameter) were exploited by Sekoguchi et al., 1984 for gas slug velocity measurements in vertical ducts. Wedin et al. (2000, 2003) used Doppler signatures to access the velocity of O(100) micrometer bubbles with a single mode fiber (core diameter 9µm): in that case, the measured bubble velocity was corrected for the deceleration experienced by the gas inclusions when approaching the probe. To investigate surface rheology, Davoust et al., 2000 exploited Doppler technique on capillary waves. Recently, Chang et al., 2003 measured the velocity of solid particles (15µm in size) transported in a water jet using a mono-mode fiber (core diameter 8µm). Later, the same sensor was used on millimeter size bubbles in homogeneous flows up to 10% in void fraction by Lim et al., 2008 as well as on large (a few millimeters) oil droplets suspended in water by Do et al., 2020. Although known for a long time, that velocity measuring technique has not become a standard, possibly because of the high frequencies, typically many MHz, to be processed. Indeed, for a $\lambda_0$=1550nm wavelength, a 10m/s air-water interface provides a 12.9 MHz Doppler frequency. Let us quote that, in the 80-90's, sensors of that type were manufactured by Nihon Kagaku Kogyo Corp. and commercialized by Kanomax Int. Corp. (Sekoguchi et al., 1984) but these products do not seem to be available any more. Nowadays, high-speed (typ. ≥500MHz) converters with large storage capability (typ. 1 Giga sample at 8bits) are commonly available (at reasonable prices) and are well suited for the Doppler technique requirements. Beside, the Doppler technique has not been attempted in complex bubbly flows such as those encountered in bubble columns operated in the heterogeneous regime where the turbulent intensity is very large (up to 70%), where bubble trajectories could take any angle with respect to the vertical, and where the flow shifts between upward to downward directions. To our knowledge, the only exception is due to Chen et al., 2003 who used the above-mentioned Kanomax product to investigate bubble dynamics in three bubble columns (0.2, 0.4 and 0.8 m I.D.) up to superficial velocities about 9cm/s i.e. within the beginning of the heterogeneous regime. These authors used a fiber with a cleaved extremity and whose diameter has been diminished from 5mm down to 350µm. They provide radial profiles of void fraction, mean bubble velocity and average bubble arrival frequency for a probe directed downward (i.e. facing the mean flow) as well as some data for a probe directed upward. Although they also provide a few bubble velocity distributions, the authors do not explain how they processed the raw signals, they do not discuss sensitivity to processing nor sensor performances and they do not explain how they built the distributions from the two data series collected with an upward and with a downward oriented probe.

*2.1 New sensor design*
We therefore revisited that technique with the objective of producing reliable data on bubble velocities in challenging flow conditions. Our bet was that the conditions for collecting the light reflected by an interface back into an optical fiber should be stringent enough to provide a way to discriminate between bubble trajectories. Moreover, we introduced an extra novelty on the sensor design to solve another issue. Indeed, all previous attempts to exploit the Doppler technique were made using cleaved tips normal to the fiber axis. Optical fibers having significant outer diameters (typically 100µm), such cleaved extremities are far from being satisfactory with respect to phase detection because of significant disturbances appearing on raw signals (Cartellier, 1990). Since our goal was to simultaneously access to the phase indicator function and to measure the velocity of bubbles, the probe tips need to be sharpened to improve the reliability of gas dwell time determination (see the discussion in Vejrazka et al., 2010). The challenge was therefore to manufacture sharp probes that provide a good contrast between water and air, and also that are able to detect Doppler signals of large enough amplitude. That objective is not straightforward. Indeed, the Doppler amplitude is maximized for cleaved tips and it is expected to become weaker when sharpening the probe tip. Meanwhile, good contrasts between air and water responses are observed for sharpened tips (Frijlink, 1987; Cartellier and Barrau, 1998a) that are also well adapted for a smooth piercing of incoming interfaces. A balance within these opposite constraints was finally reached by adjusting the curvature at the very end of conical tips (Fig.2). By fine-tuning that curvature, sharpened tips could be made as reflective as cleaved fibers, thus granting an optimum design.

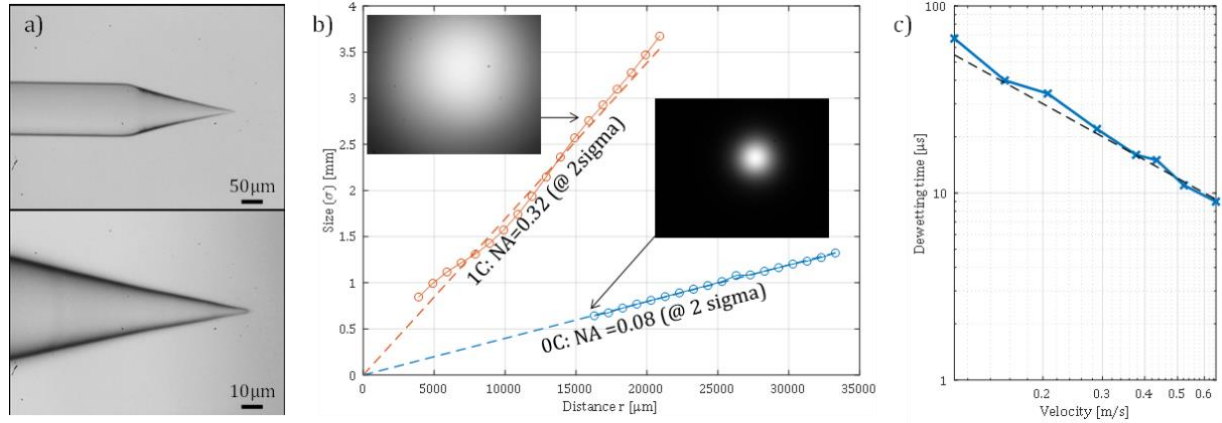

Fig.2: a) Example of a conical tip manufactured from a single mode optical fiber - the two images correspond to the same probe at different magnifications. b) Light intensity distributions measured at various distances from the fiber extremity and determination of the numerical aperture for cleaved (0C) and conical (1C) extremities. c) Measured de-wetting time $T_m$ versus the interface velocity V for the conical probe: the latency length as defined between thresholds at 10% and at 90% of the full signal amplitude is $L_S = T_m V \approx 6\mu m$. That relationship is represented by the dotted line.

The prototypes exemplified Fig.2 were manufactured by A2 Photonic Sensors using a patented technology. Corning® SMF28 single mode fibers with a 8.2µm core diameter surrounded by a 125µm cladding were used. These fibers exhibit a $n_{eff}$=1.47 effective group index of refraction at the operating wavelength. The optical fiber was fed via an optical circulator by a $\lambda_0$=1550nm laser light emitted by a 10mW distributed feedback (DFB) laser: the narrow spectral width of this device guarantees a coherence length over 5mm. The circulator output was connected to a photodetector with a 250MHz bandwidth. The optical fiber was glued into a stainless steel tube 0.9mm external diameter and 30mm long: the fiber extremity was located 7mm away from the end of the metallic tube to minimize the perturbation induced in the flow. The raw signals were digitized by a 12 bits analog-to-digital converter (ADC) before further signal processing. In order to accurately sample the high frequency Doppler bursts, the ADC must have both a high acquisition frequency (typically tens of MHz) and a large enough memory. Fig.3 provides a typical signal collected in an air-water bubbly flow with the prototype Doppler probe shown Fig.2. The signal exhibits both a clear contrast between water (≈20mV) and air (≈0.45V) and a limited noise (≈10mV in amplitude) so that the events corresponding to bubble entry and to bubble exit can be easily identified. In addition, the signal exhibits Doppler oscillations of significant amplitude both at gas-liquid and at liquid-gas transitions. The amplitude of these Doppler oscillations is significantly larger at the gas-liquid transition (from 50 to 300mV) than at the liquid-gas transition (from 20 to 90mV). That feature is due to the magnitude of the Fresnel coefficient $R_{12}$ that provides the reflected intensity with respect to the incoming intensity $I_0$: one has $R_{1/2}=[(n_1-n_2)/(n_1+n_2)]^2$ under normal incidence at the boundary between two media of refractive indices $n_1$ and $n_2$. The reflection coefficient at the fiber tip /external medium interface amounts to 3.6% in air and to 0.25% in water. The strong contrast between these two values (by a factor larger than 10) forms the basis of the phase detection capability of optical probes. Concerning the Doppler amplitude, it is related to the intensity $(I_1 I_2)^{1/2}$ of the coherent beating between the two returning waves, where $I_1$ denotes the light intensity reflected at the fiber tip (i.e. $I_1$=$R_{fiber/ext} I_0$, $I_0$ is the light intensity injected in the fiber) and $I_2$ the light intensity returning in the fiber after reflection on the moving interface. The latter wave experiences a transmission from the fiber into the external medium, (the latter is named here medium ext1, and the transmission coefficient is 1-$R_{fiber/ext1}$), a reflection at the moving interface hat separates media a1 and 2 (the coefficient is $R_{ext1/ext2}$ for a specular reflection) and a back transmission from the external medium 1 into the fiber (the transmission coefficient is 1-$R_{fiber/ext1}$), so that the returning intensity is given by $I_2/I_0$=$R_{ext1/ext2}$ (1-$R_{fiber/ext1}$)$^2$. Hence, $I_2/I_0$ has nearly the same magnitude whatever the external medium since the transmission coefficients are almost the same for air and for water. Consequently, the Doppler amplitude, that is proportional to [$R_{fiber/ext} R_{ext1/ext2}$]$^{1/2}$ (1-$R_{fiber/ext1}$) $I_0$, is controlled by the intensity reflected at the fiber tip $I_1$ that is by the reflection coefficient $R_{fiber/ext1}$. That feature explains why the amplitudes of Doppler signals are larger when the probe is in air than when it is in water. For the example of Fig.3, the ratio between the maximum Doppler amplitudes recorded at the gas-liquid transition and at the liquid-gas transition respectively is about 3.4 which is close to the expected value i.e. ($R_{fiber/air}$/$R_{fiber/water}$)$^{1/2}$=(3.6/0.25)$^{1/2}$≈3.8. This good agreement holds for an interface very close to the fiber, a position that maximizes the Doppler amplitude.

At larger distances, the spatial distribution of the illumination intensity (see Fig.4 below and the associated discussion) as well as light collection conditions must be accounted for. They both induce a decrease in the Doppler amplitude as the distance to the fiber increases, as exemplified Fig.3. Clearly, the Doppler signals extend much further ahead of the phase transition from air to water than from water to air: for a fixed Doppler amplitude about 40mV, the former extends up to 33µm while the latter extends only up to 10µm. That feature is directly connected to the magnitude of the reflection coefficient discussed above, and it is thus quite general. Globally, Doppler signals with the largest amplitude are collected either when the probe exits a gas bubble or when the probe enters a drop. These two situations also correspond to longer Doppler signals.

Going back to the example of Fig.3, eq.(1) holds for both Doppler signals so that two velocities can be estimated for the same bubble. The Doppler frequency before the liquid-gas transition is about 500kHz and it corresponds to 29cm/s. At the gas-liquid transition, the Doppler frequency is about 330kHz and it provides a velocity about 26cm/s. These velocity estimates are consistent within 10%. In this example, the second transition (i.e. from gas to liquid) corresponds to a probe leaving the bubble: it provides a slightly smaller velocity compared with the bubble entry (i.e. from liquid to gas) because of the bubble deceleration induced by the probe-bubble interaction as discussed and quantified in Vejrazka et al., 2010. The impact of the deceleration on the reliability of velocity measurements will be quantified in the next section.

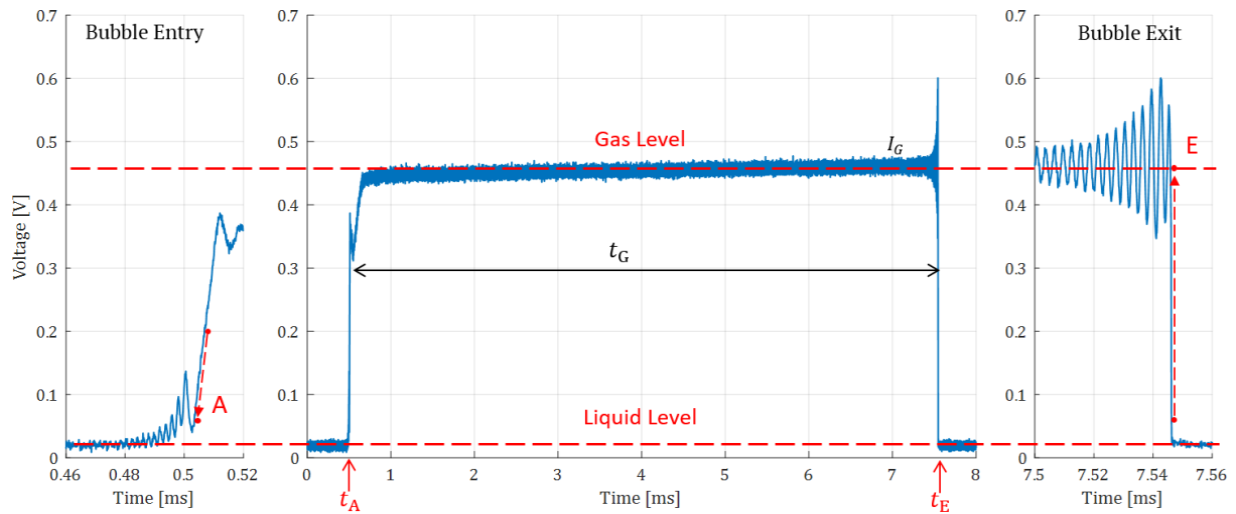

Fig.3: Typical signal from a bubble recorded with the prototype conical single mode optical probe in the bubble column described in section 3. A low voltage (here ≈ 20mV) corresponds to a probe tip in water while a high voltage (here ≈ 0.45 Volts) corresponds to a probe tip in gas. The characteristics events, i.e. arrival time $t_A$, exit time $t_E$ and the gas residence time $T_G$ are indicated on the crenel-like signal. Doppler signals collected before the probe hits the bubble (signal to the left) and when the probe leaves the bubble (signal to the right) are magnified for clarity.

We therefore succeeded to design a new sensor that provides information on the velocity of the approaching interface and that also ensures efficient phase detection. In particular, its latency length $L_s$ (see Cartellier, 1990) that quantifies the sensor resolution on the interface position determination happens to be quite small, close to 6µm (Fig.2c). Such a resolution is significantly better than that of classical multimode optical fibers for which the latency length is typically in the range 30 to 60µm. Beside, no sensor calibration is required: once the light wavelength in the propagation medium is known, the velocity is uniquely determined by the Doppler frequency using Eq.(1). A last advantage of the technique is the good reproducibility of the manufacturing process. Let us now examine in more detail the performances of that new sensor called Doppler probe in the sequel.

*2.2 Conditions for collecting a Doppler signal and consequences on the nature of the detected velocity*

As various velocities are associated with an inclusion and its interface, it is important, as we did for Laser Doppler Velocimetry (LDV in short) applied to large inclusions (Cartellier and Achard, 1985), to determine the nature of the velocity detected with the sensor developed here. Let us first underline that eq.(1) involves the *interface displacement velocity* that is, by nature, directed along the normal to the interface. This argument holds for clean interfaces that provide specular diffusion. If solid particles are attached to the interface, and if the light diffusion by these particles is intense enough to compete with specular diffusion, then the probe may also become sensitive to the velocity component of the particles tangent to the interface. Such a situation was never encountered in all the experiments we performed using unfiltered tap water, and, in the following, we do not consider nor discuss the potential impact of solid particles on the interface.

To identify the nature of the measured velocity, one should account for the conditions required for collecting a Doppler signal. The latter consists into two main constraints: first, the incoming interface must be located within the zone illuminated by the fiber, and second, the incoming interface must be oriented in such a way that the light reflected at the moving interface enters back the fiber.

2.2.1 *Illuminated zone ahead of the sensor.* That zone is partly controlled by the numerical aperture (NA) of the optical fiber. NA provides the illumination angle $\Theta$ with respect to the optical axis in an external medium of refractive index $n_{ext}$ according to $n_{ext} \sin(\Theta) = $ NA. The numerical aperture depends on the fiber design but it also changes with the shape of the tip: in particular, the NA increases for sharp tips compared with cleaved extremities. The numerical apertures were directly measured from intensity maps (Fig.2-b) recorded in air at various distances from the fiber exit. At each distance, a Gaussian normalized by the maximum in intensity at that distance was fitted, and its full-width at a given intensity threshold was plotted versus the distance to the probe to give access to the numerical aperture: this process is illustrated Fig.2-b for cleaved and for conical tips. The corresponding numerical apertures are given in Table 1 for various thresholds in intensity. For the cleaved fiber, NA defined for a 1% threshold (a 1% threshold corresponds to a width of 3 standard-deviations on a Gaussian intensity distribution) is about 0.12, which is close to the manufacturer value (0.14). For the conical tip, NA is larger: it ranges from 0.32 when defined at threshold in intensity (the latter corresponds to a width about 2 standard-deviations on a Gaussian intensity distribution), up to 0.45 for an intensity threshold set at 1%. For NA=0.45, the emission angle $\Theta$ equals 27° in air and 20° in water.

|          | Cleaved fiber (measured at 1%) NA=0.12 | Sharpened tip (measured at $1/e^2$) NA = 0.32 | Sharpened tip (measured at 1%) NA = 0.45 |
|----------|----------------------------------------|-----------------------------------------------|------------------------------------------|
| in air   | $\Theta \approx 8°$                    | $\Theta \approx 19°$                          | $\Theta \approx 27°$                     |
| in water | $\Theta \approx 6°$                    | $\Theta \approx 14°$                          | $\Theta \approx 20°$                     |

Table 1: Measured numerical apertures and corresponding illumination angles in air and in water.

These emission angles provide a rough estimate of the lateral limits of the illumination area at short distances from the fiber tip. Extra conditions must be taken into consideration to identify the active zone of the sensor. Indeed, along the axial direction, the light coming out of the fiber can be considered as a spherical wave, and its intensity strongly decreases with the distance r to the fiber extremity. Thus, any absolute detection threshold, either set by the detector characteristics or by an user-defined value, will control the maximum working distance of the probe. To evaluate that distance, the light intensity emitted out of a conical fiber tip has been simulated as a Gaussian beam with a numerical aperture given by the measured NA (consequently, the waist dimension is fixed for a given wavelength). Iso-intensity contours normalized by the maximum intensity at the waist $I_0$ are mapped in Fig.4. Strictly speaking, the results shown Fig.4 do not correspond to the actual Doppler amplitude: they account for the decay of the incoming intensity with the distance from the probe tip but they do not account for reflection and/or transmission coefficients, and they do not account for the necessary conditions for collecting some light back into the fiber. However, they remain meaningful for the present discussion because the Doppler amplitude remains proportional to the incoming intensity plotted in Fig.4. For an absolute incoming intensity equal to 1% of $I_0$, the axial extent of the probe volume in air is about 42-43μm along the optical axis. The radial extend of that zone is maximum at a distance $r \approx 25μm$ and it equals 6-7μm. Hence, the effective illumination angle $\Theta_{eff}$ as defined for an absolute intensity threshold of 1% of $I_0$ is at most about 15° in air (Fig.4 top): it is therefore significantly smaller than the illumination angle $\Theta$ deduced from the numerical aperture (see last column in Table 1). That drastic diminution of the numerical aperture is mainly due to the strong decrease of light intensity with the distance to the probe. This result also indicates that the effective illumination angle $\Theta_{eff}$ can be controlled through signal processing by setting a minimum on the Doppler amplitude. For a probe tip immersed in water (Fig.4 bottom), an intensity equal to 1% of $I_0$ is reached at a 80μm distance along the axis, and at a maximum lateral extent about 7-8μm at a 45μm distance: the corresponding effective illumination angle in water happens to be $\Theta_{eff} \approx 10°$, which is slightly less than in air for the same intensity threshold.

To consolidate the above analysis based on Gaussian optics, we measured the probe working distance by collecting raw signals from planar air-water interfaces obtained by filling or emptying a large (5cm I.D.) cylindrical tube. For these tests, the conical probe was centered in the tube, its optical axis was perpendicular to the incoming interface, and the interface motion was controlled and steady (i.e. at constant speed). To compare the signals collected at various interface velocities, the time scale was transformed into the distance r between the interface and the probe extremity using the speed of the interface as measured by the Doppler technique. The raw signals delivered by a 1C probe operated at constant source intensity and fixed detector gain, are shown Fig.5 for two interface velocities, namely 1cm/s and 59cm/s. On both signals, the working distances in air and along the optical axis are about 30μm for a threshold set at 5% of the maximum Doppler amplitude. Such distances are comparable to the ≈19μm value predicted from the emitted light intensity distributions shown Fig.4. These working distances depend on the Doppler amplitude only: they do not vary with the interface velocity. Moreover, they are not limited by light coherency since the coherence length of the diode (about 5mm) is always much larger than the optical path. Let us also point out that, for an interface travelling over a distance L, the number of successive Doppler periods equals $2\, n_{ext}\, L/\lambda_0$. This is indeed observed in Figure 5, where 26 periods are recorded over a 20µm distance in air. In water, the number of periods would have been 34.

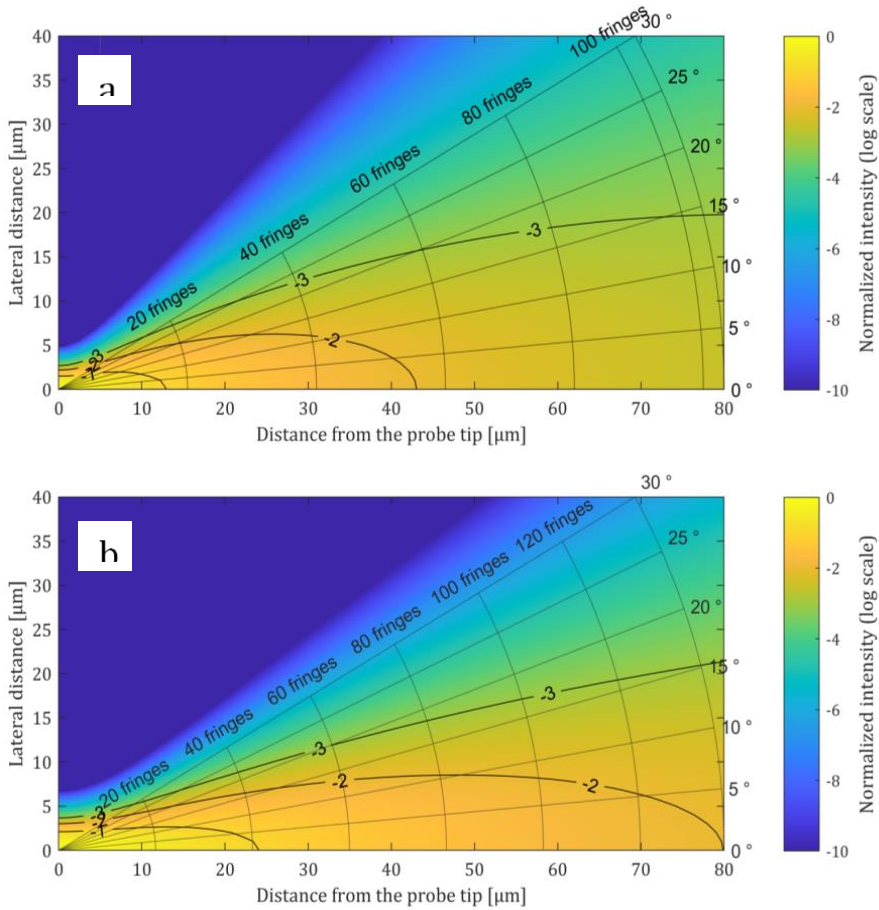

Fig.4: Light intensity I coming out of the conical fiber tip in air (a) and in water (b) normalized by the maximum light intensity at the waist $I_0$. The color represents $\log_{10}(I/I_0)$. Iso-value contours (solid lines) are provided for $I/I_0 = 10^{-1}$, $10^{-2}$ and $10^{-3}$. The number of fringes corresponds to the number of Doppler periods that would be collected for an interface travelling from the current location to the probe tip localized at the origin.

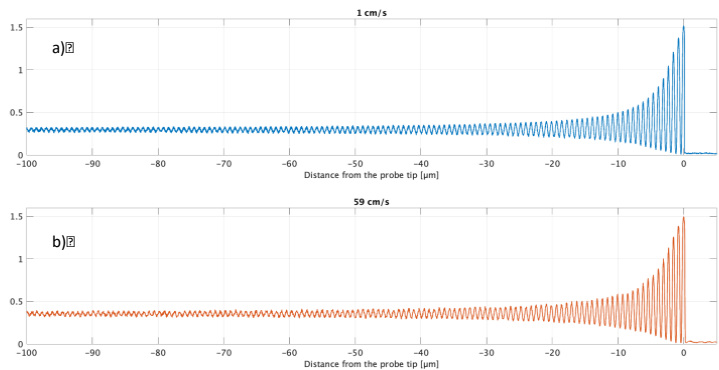

Fig.5: Signals collected from a planar air-water interface approaching a conical probe at constant speed (case a: 1cm/s; case b: 59cm/s) plotted versus the distance between the interface and the probe tip. The transition is from air to water, and the interface displacement velocity is directed along the fiber optical axis.

It is worth at this stage to compare the responses of cleaved (0C) and of conical (1C) tips. Typical signals are presented Figure 6: these signals have been simultaneously acquired from 0C and 1C probes located side-by-side and facing the same planar interface. The optical axes of both probes were normal to the interface. In addition, comparable source intensity and detector gain were used on the two probes. Clearly, the cleaved probe has a much longer working distance than the conical probe as it is indeed able to detect Doppler signals up to a millimeter ahead of its tip compared with 50-100µm for the conical probe. This is because the smaller numerical aperture of the cleaved fiber induces a smaller decay rate of the incoming intensity with the distance to the probe. This is also because the aperture available to collect the reflected light is much larger (by a factor close to 10) for the cleaved fiber compared with the conical tip. Consequently, the number of recorded Doppler periods is typically 100 times larger in the latter case. Thus, and compared with cleaved fibers, conical probes provide a local information. In addition, conical probes induce much weaker interface deformation prior to piercing: this is visible on Fig.6 bottom where the measured interface velocity remains almost constant till contact for the conical probe while, for the cleaved fiber, the velocity strongly decreases 5 to 10µm before contact. In addition, cleaved tips have been early abandoned in the development of phase detection sensors (Cartellier and Achard, 1991) because of the various defects they induce including local interface deformations during piercing that produce distorted signals hard to process, but also alterations of the bubble trajectory with subsequent errors on detected chords distribution. For all these reasons, we selected conical fiber tips to develop further the Doppler technique for dispersed two-phase flows.

The sensor geometry being set, the discussion on the nature of the detected velocity can be pursued. The iso-contours presented Fig.4 define the region into which the interface must be located to collect a Doppler signal of given amplitude. Whatever the threshold selected between 0.1% and 10% of $I_0$, the volume it defines has a quite limited extent, of the order of a few tens of micrometers: that confirms that the velocity information arises from a very small region on the interface of the approaching inclusion.

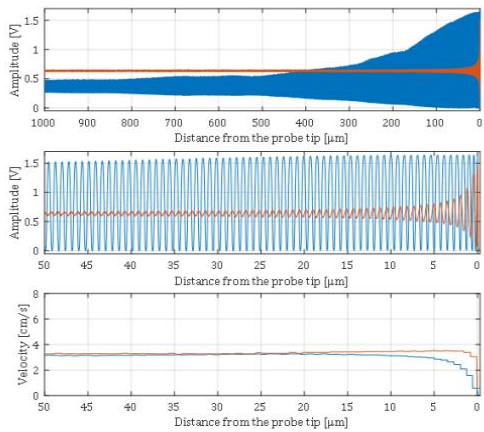

Fig.6: Signals from a cleaved probe (blue signal) and from a conical probe (orange signal) simultaneously collected from the same planar air-water interface: top figures correspond to the same signal with different magnifications. The bottom figure provides the interface displacement velocity versus time as measured from the Doppler signal delivered by each probe.

2.2.2 *Interface orientation and conditions for collecting some reflected light*: Let us discuss now the second constraint required for forming a Doppler signal, i.e. the collection of reflected light back within the fiber. This requirement induces stringent angular conditions. On one hand, the numerical aperture provides the maximum angle possible for collecting external light into the fiber. In addition, the light reflected by the incoming interface must re-enter the fiber by hitting the fiber core at its very tip. The latter is quite small: for the single-mode fibers used here, the core diameter d is 8.2 µm and the tip size $d_c$ is even smaller due to fiber sharpening. Hence, as shown Fig.7-a, for an approaching interface along any incident vector **k**, the returning ray must be at an angle less than $\beta/2$ with $\tan(\beta/2)=d_c/(2r)$ with respect to **k**. That means that the local normal to the interface **n** must be at most at an angle $\pm\beta/2$ with **k**. Figure 7b provides the magnitude of $\beta/2$ versus the distance r for various tip sizes, namely $d_c$=2, 5 and 8µm. For estimating the angle $\beta$, the conical shape of the fiber extremity was not accounted for, and we considered instead a circular collecting area of diameter $d_c$. The resulting angle $\beta/2$ remains quite small (less than 10°) at "large" distances from the fiber tip (say above 20-30µm). It neatly increases when the reflector gets closer to the fiber, notably over a range of distances where the Doppler amplitude becomes large (see Fig.5). However, $\beta/2$ cannot exceed $\Theta$ as set by the numerical aperture since that angle corresponds to the largest incidence on the fiber tip that allows the transmission of light back into the fiber. Hence, the condition for collecting some reflected light within the fiber writes $\mathbf{k}\cdot\mathbf{n} \leq \cos[\text{Min}(\pm\beta/2, \Theta)]$.

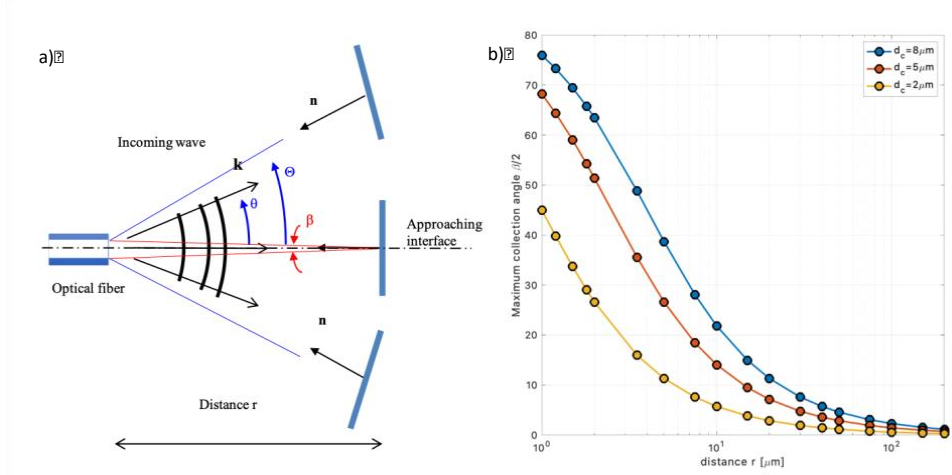

Fig.7: a) Schematized emission from the extremity of an optical fiber and angular conditions for detecting an approaching interface. b) Maximum collection angle β/2 allowed for forming a Doppler signal as a function of the distance r for various diameters $d_c$ of the probe extremity (the latter is approximated as a flat circular area).

The above reasoning is strictly valid for a vector **k** directed along the optical axis (i.e. θ=0). For a wave vector at an angle θ with the optical axis (with θ in the interval ± Θ), the incidence of the returning wave on the fiber tip must be within the collection angle that leads to the condition θ±β/2 less than Θ. Therefore, as θ increases, the range of β angles leading to a Doppler signal narrows, and in the limit θ=Θ, the only active ray left corresponds to the incident vector **k**. Overall, for Doppler signals to appear, the maximum inclination of the normal **n** to the incoming interface with the optical axis should be ±Θ. This result and the above discussion have two consequences:

- first, the maximum inclination β/2 of the interface displacement velocity with respect to the active wave vector is ±Θ. Therefore, a first source of uncertainty on the quantification of the displacement velocity arises because of the range of eligible interface inclinations for a fixed vector **k**. In other words, the angle between **n** and **k** involved in eq.(1) varies from 0 to Θ, so that the corresponding uncertainty on the velocity magnitude amounts for cos(Θ) at most. Note that this is a somewhat conservative estimate as the range of β angles could be significantly less than Θ for very thin tips (i.e. small $d_c$) as shown Fig.7-b.

- second, the active vector **k** can vary when the interface approaches the fiber tip, and one has to account for such a change during the formation of a Doppler signal. To discuss this aspect, let us introduce the interface displacement velocity projected along the optical axis, noted $V_{axis}$. According to the above discussion, the velocity detected using eq.(1) can evolve between $V_{axis}$ (for an active wave vector such that θ=0) and $V_{axis}$ cos(θ) cos(θ−β/2) or $V_{axis}$ cos(θ) cos(θ+β/2) when changing the wave vector within acceptable limits. Taking the latter as ±Θ, and since the condition θ±β/2<Θ always holds, the maximum deviation from $V_{axis}$ is therefore $V_{axis}$ $cos^2(Θ)$. A more realistic estimate would consider that the acceptable limits for the inclination of the wave vector correspond to the effective illumination angle $Θ_{eff}$ introduced above, in which case the maximum deviation from $V_{axis}$ amounts to $V_{axis}$ cos(Θ) cos($Θ_{eff}$).

Overall, the viewing angle (possibly combined with its effective value) controls the uncertainty on velocity when the latter is interpreted in terms of the interface displacement velocity projected along the probe optical axis. Fig.8 provides the magnitude of that uncertainty versus the numerical aperture, either using the conservative estimate $cos^2(Θ)$, or using the more realistic estimate cos(Θ) cos($Θ_{eff}$). According to Fig.8, the uncertainty on velocity increases with the numerical aperture. In term of sensor optimization, cleaved fibers provide small apertures (Table 1) but, and as already argued, such geometries are not suited for phase detection. Instead, conical tips should be preferred. For the conical sensors we manufactured, NA ranges from 0.32 to 0.45 (depending on the threshold considered, see Table 1), and thus the typical uncertainties range from 10 to 20% in air, and from 6 to 12% in water. Using the more realistic estimate related with the effective aperture $Θ_{eff}$ based on the $10^{-2}$ iso-intensity contour, these uncertainties evolve from 8.4% to 13.7% in air and from 4.3% to 7.2% in water. Overall, the uncertainties are less than ≈14% for a probe in air, and less than ≈7% for a probe in water: they are thus acceptable. It is worth underlining that the effective angle, and thus the uncertainty, could be further reduced by selecting Doppler signals of larger amplitude, but possibly at the price of lower data rates.

According to the conditions required for collecting a Doppler signal, it happens that the proposed sensor provides the *interface displacement velocity projected along the probe optical axis*. In addition, the smaller the numerical aperture, the smaller the uncertainty on the measured velocity magnitude.

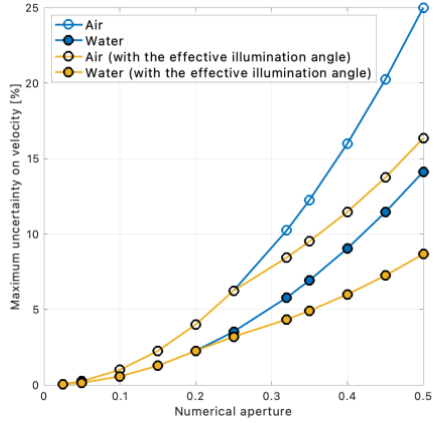

Fig.8: Maximum uncertainty (in percent) on the magnitude of the interface displacement velocity projected along the probe axis as a function of the numerical aperture for a probe in air (open symbols) or in water (closed symbols). Results are provided both for the nominal and for the effective numerical aperture.

The companion question to be considered now is to connect the information on the interface displacement velocity with the translation velocity of the inclusion, a key variable for flow analysis and modeling. We have seen that the volume probed by that technique is quite small – typically O(10µm) so that only a very limited portion of the interface of incoming inclusions is scanned by the technique. We have also seen that, to observe Doppler signals, the normal of incoming interface must make an angle within ±Min($\Theta$, $\Theta_{eff}$) with the probe optical axis, where $\Theta_{eff}$ depends on the selected amplitude threshold on Doppler signals. For an air-water interface, and for a threshold $10^{-2}$ $I_0$, $\Theta_{eff}$ has been shown to be about 15° in air and 10° in water. Consequently, the velocity information always arises from the close vicinity of the apex of the front (or rear) interface of bubbles. For spheres, for ellipsoids as well as for distorted, wobbling bubbles (Clift et al., 1978), the interface velocity in these zones is very close to the translation velocity of the inclusion: the difference is driven by the cosine of the angle and it is less than 3% in water and less than 5% in air. Hence, one can conclude that the proposed sensor gives access to the *translation velocity of the inclusion projected along the probe optical axis*.

A similar reasoning has been previously used to interpret the velocities deduced from the de-wetting time as measured from multimode optical probes. The angular conditions required to collect velocity information with the new sensor are more stringent (±10° in water, ±15° in air) than those for multimode optical probes. Indeed, for the latter that are based on the de-wetting time/velocity relationship, the calibration remains valid within 10% up to an incidence angle about 20° for conical probes (Cartellier, 1992; Cartellier and Barrau, 1998a) and about 40° for 3C (i.e. conical-cylindrical-conical) probes (Cartellier and Barrau, 1998b). Above these angles, the rise time sharply increases with the incidence and it is no longer possible to reliably evaluate the velocity. For the Doppler technique, the angular conditions of impact between the inclusion and the probe do not affect the velocity measurements because it is an absolute method, and there is no uncertainty or bias on the measured velocity due to the incidence. The only limitation arises from the fact that the Doppler frequency is sensitive to radial interface motion, so that the velocity measurement could be perturbed by local oscillations of the surface or by the radial expansion (or deflation) of the inclusion. Indeed, an inclusion whose radius a is changing in time (increasing or decreasing) at a rate da/dt would generate a Doppler signal with a frequency equals to $2|da/dt|/(\lambda_0/n_{ext})$. Therefore, the Doppler frequency recorded with the proposed sensor is sensitive both to the translation velocity of the inclusion U and to its radial motion da/dt, and there is no easy way to disentangle these two contributions. Yet, experimental situations for which da/dt reaches values comparable to that of the translation velocity are not common: that can possibly occur during phase change (cavitation, boiling, condensation) or with forced interface oscillations at high frequency (e.g. due to acoustic forcing) or during rapid interfacial phenomena such as ligament pinch off, recession of a thin rim, film rupture during coalescence... Whenever such quick interface deformations are absent, the Doppler frequency as measured with the new sensor provides the translation velocity of the inclusion projected along the probe axis.

Although we have seen that water-to-air transitions are advantageous in terms of velocity uncertainty, air-to-water transitions will be considered from now on because that situation provides the largest amplitude for Doppler signals.

2.2.3 *Sensitivity to inclusions trajectory*

To exploit the proposed technique in bubble columns operated in the heterogeneous regime, it is important to determine how the sensor responds to inclusion trajectory, and in particular to inclusions coming from the side or from the rear. For that, we performed controlled experiments in which the incidence angle α between the probe axis and the normal to the planar interface was varied. These tests were achieved on the same installation as before by tilting the tube (with the probe, still aligned along the tube axis and tilted as well). The corresponding raw signals gathered at air to water transitions are given Fig.9.

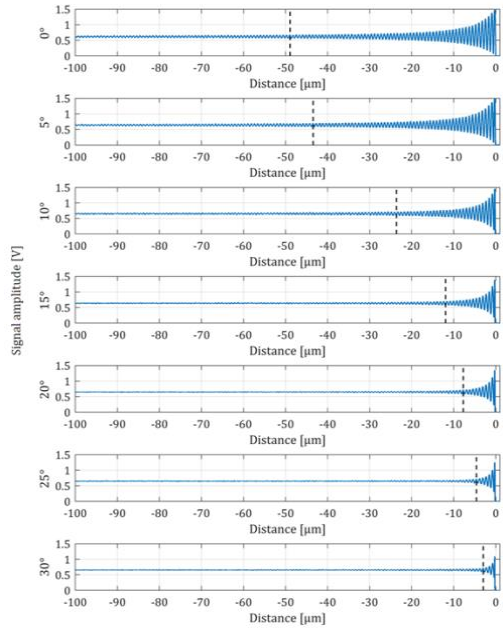

Fig.9: Raw signals occurring at air to water transitions collected with the same mono-mode conical probe for different angles α between the normal to the interface and the probe axis. The signals in volts versus time have been transformed into volts versus distance to the probe tip using the velocity measured with the Doppler frequency. The dashed lines correspond to a fixed signal amplitude equals to the static gas level plus 9 (±1) times the noise standard deviation.

All signals in Fig.9 were acquired with the same source intensity, the same amplification at detection and with identical digitalization conditions: they can thus be compared in terms of amplitude and duration. To analyze these signals, we considered the interface position such that the Doppler amplitude reaches a fixed value, equal to the static gas level (i.e. the voltage output of the sensor when its tip is in the air) plus 9(±1) standard deviations of the signal noise distribution. For each incidence angle, the corresponding position of the interface is represented by a dash-line in Fig.9. Its distance L to the probe tip is plotted versus the incidence angle in Fig.10 together with the number of successive periods (#) registered over that distance (the number of periods equals $2n_{ext}L/\lambda_0$ as expected). Fig.10 also provides the maximum Doppler amplitude recorded for an interface almost touching the probe as a function of α. The maximum Doppler amplitude slightly diminishes with α, while the distance L to the interface corresponding to a fixed Doppler amplitude sharply decreases when the inclination exceeds 5°. In other words, the signal neatly shortens as the incidence increases above ≈10°, and the signal disappears for an incidence above 30°. Hence, if the minimum number of Doppler periods is set to 20 in the processing criteria, then no Doppler signal is validated when the incidence α exceeds 12-13°. Consequently, we recover an effective collection angle smaller than 15°, and thus the maximum uncertainty on velocity measurements remains below 14% as discussed in the previous section. A more stringent criterion on the minimum number of Doppler periods present in the signal would restrict even further the range of eligible interface inclinations with respect to the optical axis. In addition, such a criteria induces also a limit on the maximum inclination of inclusion trajectories.

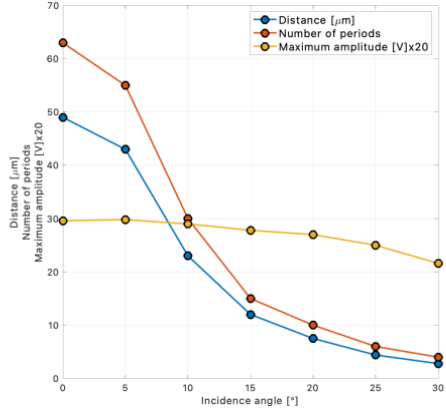

Fig.10: Evolution with the angle α between the normal to the interface and the probe axis, of the maximum Doppler amplitude (in Volts) reached when the interface almost touches the probe, of the number of successive Doppler periods in the signal (#) and of the distance (in µm) for which the Doppler amplitude remains higher than 9(±1) times the standard deviation of noise distribution.

Indeed, the results from Fig.9 and 10 can be exploited further to identify eligible bubble trajectories. For that, the number of Doppler periods recorded as a function of the interface inclination has been represented as a number of fringes versus the angle α on the intensity map of Fig.4: Fig.11 indicates that the resulting boundary is rather close to the $10^{-2}$ iso-contour in intensity. That boundary controls the trajectories compatible with a given maximum extend of Doppler signals, or equivalently with a given number of periods. Clearly, from Fig.11, it happens that only an inclusion travelling along the optical axis can produce a signal 60 periods long. If one considers now shorter signals that are at least says n periods long, eligible trajectories are such that i) they are fully within the dashed line iso-contour (that sets a minimum amplitude for Doppler signals) and ii) their projection along the optical axis (here the horizontal) is at least n fringes long: the example shown Fig.11 corresponds to signals of at least 20 periods. Note that the projection is required because the velocity component normal to the optical axis does not contribute to the Doppler shift. The above two conditions provide an upper bound for the inclination of inclusion trajectories with respect to the optical axis: the maximum inclination is plotted versus the number of Doppler periods in Fig.12 (for the selected Doppler amplitude). Clearly, as the minimum number of required periods increases, eligible bubble trajectories become more and more aligned along the optical axis. That trajectory selection has no consequence on measurement uncertainty because, as already discussed, the latter is controlled by the inclination of the interface and by the orientation of the active wave vector. Yet, that trajectory selection may have some consequences on the velocity statistics. Indeed, according to Fig.12, if one sets a minimum of 10 fringes, the trajectory of the inclusion must make an angle within ±44° with the optical axis. For a minimum of 20 fringes, that angular range decreases to ±28°. Similar effects are present in LDV, and as for LDV, we will show in section 3.3 that velocity pdfs as measured by the new probe are almost insensitive to signal processing parameters when the latter are properly selected.

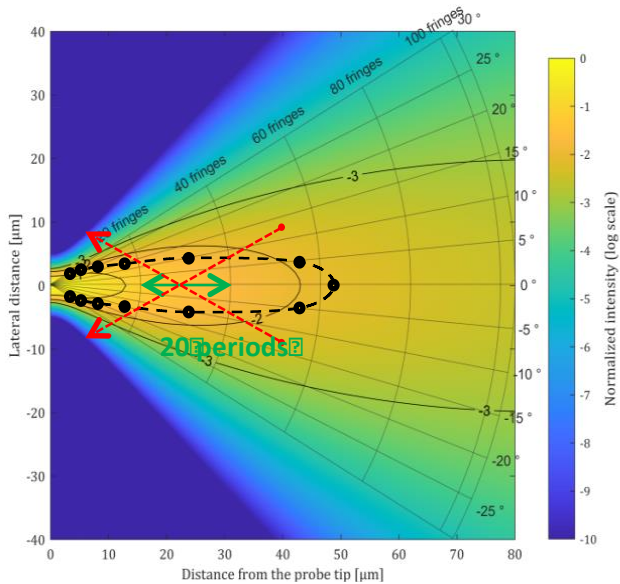

Fig.11: On the intensity map of Fig.4, the open dots represent the number of fringes (or Doppler periods) measured in controlled experiments when varying the angle α at fixed amplitude of the Doppler signals. The red dashed lines correspond to extreme bubble trajectories for a number of periods equal to 20.

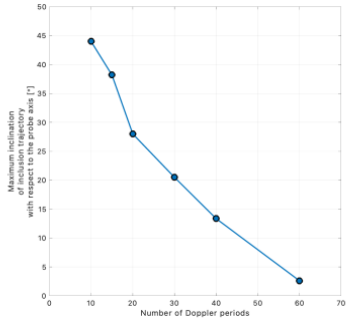

Fig.12: Maximum inclination of eligible inclusion trajectories with respect to the optical axis as a function of the number of periods recorded at an air to water transition. Here, the Doppler periods are considered only when the Doppler amplitude exceeds a given threshold.

Another important issue related with velocity measurements concerns the ability of the sensor to detect positive (i.e. corresponding to an inclusion approaching the fiber tip on its front) or negative (i.e. corresponding to an inclusion approaching the fiber tip from the rear of the probe) velocities. This question is relevant since flow reversal is commonly encountered in bubble columns operated in the heterogeneous regime. That feature has prompted users to reconstruct pdfs from two data series recorded at the same position in the flow, one acquired for a sensor pointing upward and another one from a sensor pointing downward (Xue et al. 2003, 2008; Chen et al. 2003). For these pdf reconstructions, it is usually argued that local probes do not detect gas inclusions arriving from behind because of the presence of a (thick) probe holder. Yet, experimental evidences of that behavior are scarce. In that perspective, we examined whether or not bubbles arriving from behind can produce exploitable and meaningful Doppler signatures. A companion objective was to quantify a possible cross-talk error in the signal interpretation. Two series of experiments were undertaken in that perspective.

A first series was achieved on quasi-planar interfaces, using gas slugs that were long enough to ensure the complete de-wetting of the probe holder. Fig.13 provides the signals recorded by a downward directed probe interacting with an ascending interface: the signals shown in the top row (cases a and b) correspond to the standard situation of a probe impacting a bubble approaching head on. In the bottom row, Fig.13 provides signals collected by the same probe but directed upward when an ascending slug goes through it: that inverted situation corresponds to a bubble approaching the probe from the rear (signals c and d). Let us compare the signals collected in these two situations. At liquid-to-gas transitions (left column in Fig.13), a weak Doppler signal is present before the transition in the standard situation (case a). When the probe direction is inverted, a Doppler signal still occurs at the liquid-to-gas transition but it is now located *after* the phase transition (case c). Concerning the passage from gas to liquid, a Doppler signal occurs before the transition in the standard situation (case b) but none is recorded by an upward directed probe (case d). Hence, for bubbles larger than the tube holding the probe that are able to de-wet the probe support, there is no risk to misinterpret the situation since the processing will select Doppler signals arising at the gas level and preceding the gas-to-liquid transition (i.e. the signals corresponding to case b in Fig.13).

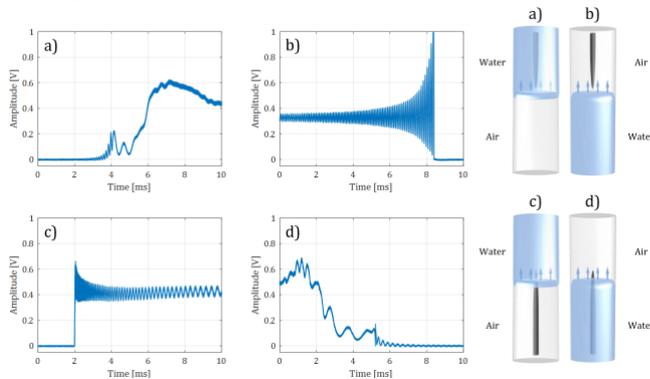

Fig.13: Probe response when crossing an interface from air-to-water or from water-to-air with a probe oriented along or against the gas slug velocity: a) and b) situations correspond respectively to bubble entry and bubble exit for a bubble approaching the probe tip head-on. Situation c) - respectively situation d) - corresponds to a probe tip entering - respectively exiting - a bubble approaching the probe tip from its rear.

A second series of experiments took place in a quasi-unidirectional bubbly flow produced in a 50mm I.D. tube. Velocities were in the range 0.5-1m/s and bubble were 1 to 4 millimeters in size. Such dimensions are much smaller than those of the probe holder: the full de-wetting of the probe support was thus unlikely. Velocity statistics were gathered for a probe facing the main flow and for the same probe oriented at 180° from the main flow direction. In the latter case, no Doppler signal was detected before gas-to-liquid transitions: consequently not a single velocity data was acquired when the probe was at 180° from the main flow. To be very precise, let us underline that a few Doppler signals were detected in that probe position, but they all occurred after the gas-to-liquid transition and thus, they were not accounted for by the processing. Hence, whatever the bubble size compared with the probe holder dimensions, the new sensor detects positive velocities only, where the term positive corresponds here to an inclusion approaching the fiber and those direction goes from the probe tip to the probe holder: this situation corresponds to the situation a) in Fig.13.

*2.2.4 Probe-bubble interaction impact on velocity measurements*

Finally, as the rear air/liquid interface of the bubble has been selected to analyze Doppler signals, it is worth evaluating the impact of the probe-bubble interaction on the change in the bubble velocity. Probe-bubble interaction has been analyzed for conical fibers used for phase detection (Vejrazka et al., 2010). The resulting uncertainty on chord and concentration was shown to be controlled by the modified Weber number $M=[\rho_L D_b u_b^2/\sigma] (D_b/d_{probe})$ where $u_b$ is the bubble approaching velocity with respect to the probe, $d_{probe}$ the outer fiber diameter, $D_b$ the bubble diameter, $\rho_L$ the liquid density and $\sigma$ the air-water surface tension (Vejrazka et al., 2010). Typically, the uncertainty on void fraction remains below 10% for M higher than 50. Such a condition is easily fulfilled in industrial bubble columns as these are usually operated at large superficial velocities so that the axial fluid velocities exceed ≈0.2m/s over most of the column cross-section. In addition, the bubbles produced by industrial spargers are typically larger than 2mm in diameter (Kantarci et al., 2005; Besagni et al., 2018; Chaumat et al., 2007) so that their relative velocity is at least 0.2 m/s in fluids with a viscosity equal to or less than that of water. These conditions correspond to bubble velocities with respect to the fixed probe of at least 0.4m/s and thus to M higher than 50. Note that such conditions hold in the 0.4m I.D. bubble column exploited in section 4.

The modified Weber number M is also a key parameter controlling the bubble deceleration during its interaction with the probe. More precisely, the change in the bubble velocity $\Delta u_b$ at the time when the probe has already pierced the bubble front interface and hits the bubble rear interface, compared with the undisturbed bubble velocity $u_b$ was shown to be $\Delta u_b/u_b \approx - 6 / [C_{AM} \chi^{2/3} M]$, where the bubble aspect ratio $\chi$ is defined as the major axis divided by the minor axis, and where $C_{AM}$ is the added mass coefficient (Vejrazka et al., 2010). Note that $\Delta u_b$ diminishes with the aspect ratio because of the strong decrease of $C_{AM}$ with $\chi$. In air-water systems, $\Delta u_b/u_b$ remains below 15% provided that the quantity $C_{AM} \chi^{2/3} M$ exceeds 40. That condition is indeed fulfilled in bubble columns because M equals 50 (at least) as discussed above, and also because $\chi$ is larger than 1.2 as soon as bubbles exceed 1.1-1.2 mm in size. Overall, the impact of the probe on the bubble slow down is at most -15%. This estimation should be revisited for fluids more viscous than water.

To summarize the analyses presented in this section, the proposed sensor ensures accurate phase detection due to its conical tip and its quite small latency length. It also provides information on velocity derived from the Doppler signals formed at the gas-liquid interface and occurring just before the sensor extremity exits the bubble. The measured velocity corresponds to the translation velocity of the gas inclusion projected along the probe optical axis. Moreover, as bubbles arriving from the rear are easily discarded by the signal processing, only positive velocities that correspond to inclusions approaching the front of the sensor are detected. The overall uncertainty on velocity measurement has been evaluated to be at most 15% when considering both optical detection conditions and probe-bubble interactions. For the latter, our estimate is grounded on a carrier phase velocity of 0.2m/s: larger liquid velocities are usually encountered in bubble columns so that the bubble deceleration should be significantly diminished. For example, for a carrier phase velocity of 0.4m/s, the modified Weber number M is doubled and the change in bubble velocity decreases down to -8% instead of -15%. Concerning the optical conditions, we have shown that imposing a higher threshold on the Doppler amplitude or adding a requirement on the minimum number of Doppler periods changes the optical constraints and diminishes further the uncertainty on velocity. On these bases, a signal processing routine has been developed to extract the variables of interest. This processing is presented in the next section together with the sensitivity of measured variables to the validation criteria introduced in that processing.

**3. Signal processing, sensitivity to processing parameters and sensor performances**

To develop the processing and to test the performances of the sensor, data were acquired in bubbly flows produced in a 3m high bubble column with an internal diameter D=2R=0.4m. The column was filled with tap water at an initial height of 2.02m. Dried air was injected through 352 orifices (1mm in diameter, 10mm long) uniformly distributed over the cross-section S. The injected gas flow rate $Q_G$ was varied from 100 to 1900 Nl/min, so that the superficial velocity $Vsg=Q_G/(\pi R^2)$ ranged from 1.3cm/s to 35cm/s. All data were collected at H=1.45m above injection, i.e. for H/D=3.62 a position within the quasi fully developed region (Maximiano Raimundo et al., 2019). Unless otherwise stated, the Doppler probe was vertical and downward oriented. The probe was immerged in the bubble column from the top using a metallic square bar (side 20mm) to avoid vibrations: to reduce the flow perturbation, the probe was located 55mm away from that bar and parallel to it.

As for classical optical probes, the signal delivered by the Doppler probe consists in a succession of crenel-like modulations, each one corresponding to a bubble pierced by the sensor (Fig.14). The goal of the processing is to determine the arrival time $t_A$, the exit time $t_E$ and thus the gas dwell time $T_G = t_E-t_A$ of each detected bubble (Fig.3) and to evaluate its velocity from the Doppler modulation when the latter is present.

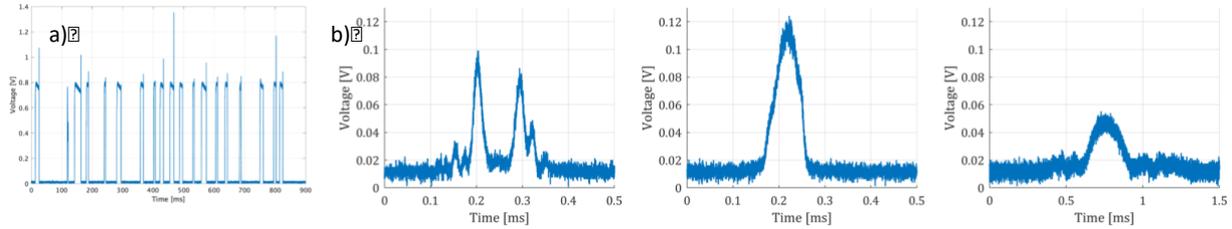

Fig.14: Raw signals from the Doppler probe (a) and examples of partial amplitude signals (b).

*3.1 Phase detection routine*

To identify the bubbles hitting the probe, we adapted the signal processing developed for classical optical probes and already tested in diverse flow conditions (Barrau et al., 1999). First, as the liquid level is quite stable, any signal exceeding the noise at the liquid level is considered to be a contribution from a gas inclusion. Accordingly, the detection routine uses an absolute threshold in amplitude: this lower threshold is typically set at the liquid level plus half the peak-to-peak noise amplitude. Second, as Doppler modulations must not be interpreted as gas inclusions, we introduced an upper threshold in amplitude that is higher than the lower threshold (Fig.3). As seen in section 2.1, the amplitude dynamics of the signal is set by the Fresnel coefficient times $I_0$. The liquid level $A_L$ corresponds to a fully wetted probe for which $R_{fiber/water} \approx 0.0025$. The plateau at the gas level $A_G$ is obtained when the entire latency length has been dried off: it corresponds to $R_{fiber/air} \approx 0.036$. Thus the signal dynamics $\Delta A=A_G-A_L$ corresponds to 0.033 $I_0$. Meanwhile, the amplitude of a Doppler signal at a water-to-air transition amounts to 0.007 $I_0$ that is roughly to 21% of $\Delta A$. At an air-to-water transition, the Doppler amplitude is larger: it amounts to 0.026 $I_0$ that is about 77.6% of $\Delta A$. These estimates are optimistic as we assume here normal incidences, and we do not account for any geometrical factor connected with light diffusion on the interface or with light collection back in the fiber. Actual Doppler amplitudes are therefore expected to be significantly smaller than the above figures. Concerning bubble detection, when a Doppler occurs before the water-to-air transition as in Fig.3-left, the amplitude of the signal is at most $A_L + (0.21/2) \Delta A= A_L+0.1 \Delta A$. Thus, setting an upper threshold above that limit ensures the detection of the gas inclusion without interpreting Doppler oscillations as bubbles. The arrival time $t_A$ of the gas inclusion is then found by going back in time to the first event whose amplitude is close to the lower threshold: this process is represented by the red arrow leading to the event A in Fig.3-left.

At the gas-to-liquid transitions (Fig.3-right), the Doppler amplitude is stronger (≈78% of the signal dynamics) but it never reaches the liquid level. Therefore, any threshold below $A_L + (1-0.78/2) \Delta A = A_L + 0.61 \Delta A$ allows detecting the gas-to-liquid transition while escaping the Doppler signal. Consequently, either an upper below $A_L+0.61\Delta A$ or the lower threshold as defined above can be used to capture the sharp gas-to-liquid transition. To identify the time $t_E$ corresponding to bubble exit, one has to go back in time to the first event those amplitude reaches the gas plateau: the red arrow leading to the event E in Fig.3-right represents that search. Hence, Doppler oscillations are not accounted for in gas detection, and the gas dwell time $t_{Gi}$ of the i$^{th}$ bubble is $t_E$ - $t_A$.

The above phase detection routine is well adapted to signals with a plateau at the gas level such as the one shown Fig.3. As the latency length of the Doppler probe is very small compared with usual bubble sizes (the latter are typically above 0.5mm in most industrial processes), nearly all bubble signatures do reach a plateau at the gas level. From a long record collected on the column axis at Vsg=3.3cm/s, we indeed observed that 98.6% of the bubbles signatures reach the plateau at the gas level. Yet, not all these signals provide a Doppler modulation: this is because of the angular constraints discussed in section 2.2. In addition, the 1.4% remaining events collected by the probe corresponds to signals of very small amplitude: examples are provided Fig.14-right. Such events occur when the probe extremity is not fully de-wetted during the passage of a gas inclusion. In the flows considered here, the bubble size is always much larger than the latency length. Therefore, all signatures of reduced amplitude correspond to tangential hits, i.e. to very small chords, of the order of the latency length, cut through larger inclusions (Vejrazka et al., 2010; Cartellier, 1992). For such signatures, the gas entry is detected as described above. For the gas exit, since these signatures are most of the time smoothly decaying in time after having reach their peak amplitude, the date of the peak is usually considered as the gas exit. For the Doppler optical probe, these low amplitude signals are rare and they bring a negligible contribution to the void fraction. At Vsg=3.3cm/s, they amount for 9 $10^{-5}$ in absolute value to be compared with the mean void fraction of 0.15. In addition, these small signatures, whose amplitude is at most 30-40% of the signal dynamics, never provide Doppler signals. Therefore, only the fraction of these small signals whose maximum amplitude is below the upper threshold is discarded by the processing.

*3.2 Phase detection efficiency*

The sensitivity to threshold selection was analyzed on a signal collected on the column axis at Vsg=19.5cm/s, i.e. in the heterogeneous regime. The 348 seconds long record (more than 40000 bubbles detected) ensures the convergence of the void fraction within ±0.0044 of void for a mean void fraction equal to 0.335. The liquid level was 7mV with a peak-to-peak noise about 23mV: the lower threshold was set at 25mW, i.e. slightly above the minimum recommended of 18.5mV. The gas level was 0.9 Volts just after piercing and it smoothly decreased in time down to its plateau value at about 0.8 Volts (see Fig.15) because of the thinning of the liquid film attached to the probe tip. Hence, according to the recommendation given in the previous section, the upper threshold should remain below 0.55mV. However, since the signals experienced strong overshoots (up to 1.5V) at the gas-liquid transition, we extended the sensitivity analysis outside the recommended range, and the upper threshold was varied from 26mV up to 1.5V. The analysis was repeated for a lower threshold twice as large (50mV) and for upper thresholds between 100mV and 1.5V. The evolution of the local void fraction versus the two thresholds presented Fig.15 shows that the void fraction remains almost insensitive to the threshold selection. Quantitatively, the maximum difference amounts to $1.5 \cdot 10^{-3}$ in absolute void fraction (less than 0.4% in relative value) when the upper threshold evolves within the recommended range. That result also holds for an upper threshold up to 1.2V: this is because the detection criterion is not absolute but it is conditional. Indeed, whenever a bubble signature is detected, the processing goes back to the entry point A defined by the lower threshold (Fig.3). The exact value of the upper threshold is therefore irrelevant provided that the presence of a bubble signature has been detected. This is no longer true above 1.2V because not all overshoots reach such a voltage, and this is why the void fraction and the number of detected bubbles decrease for an upper threshold between 1.2V and 1.5V. Similar results are obtained when doubling the lower threshold (i.e. 50mV): this is because, for the signals considered, the peak-to-peak noise remains stronger than the Doppler amplitude at water-air interfaces. Overall, the void fraction and the number of detected bubbles happen to be almost insensitive to the upper threshold when the latter is selected within a wide interval, ranging from 10% to 60% of the signal dynamics. Moreover, we also found that the gas residence time distribution remains the same when the thresholds evolve within the recommended range. We further compare with the void fraction on the column axis detected by a classical conical probe those latency length was about 40µm: from Vsg=0.3 to 25cm/s, the gas hold-up difference in relative value between the two sensors was in average 3% with a maximum at 6%. Hence, according to these tests, the Doppler sensor happens to be quite reliable in terms of gas phase detection.

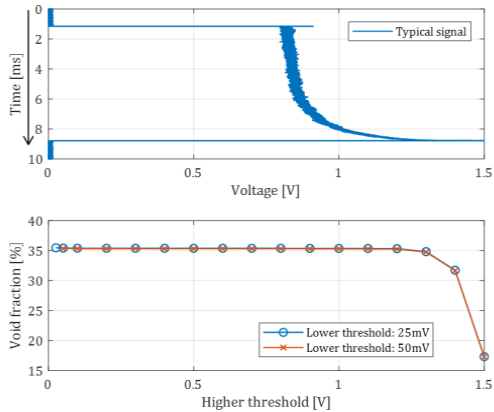

Fig.15: Typical bubble signature (top) and evolution of the void fraction versus the upper threshold for two lower thresholds (bottom). Signal 348 seconds long collected on the bubble column axis for Vsg=19.5 cm/s.

*3.3 Velocity measurement*

Once individual bubbles are identified, Doppler bursts are searched before each phase transition. That search is performed over a time window evaluated from an a priori estimate of the minimum expected velocity and for a given number of periods. The extent of the search window is also bounded by the previous phase transition. The raw signals are digitalized at a sampling frequency $f_{sampling}$ at least twice (owing to Nyquist criteria) the maximum Doppler frequency $f_{Dmax}$ corresponding to the largest velocity in the flow. In practice, $f_{sampling}$ is often set to five times $f_{Dmax}$. The default value for the minimum velocity is set to $f_{sampling}/500$: it defines the lowest Doppler frequency $f_{Dmin}$ that can be detected. A Fourier transform is then performed over the working window to identify potential Doppler bursts within the interval [$f_{Dmin}$, $f_{Dmax}$]. For that, use is made a Butterworth filter of order 2 those

-3dB bandwidth is [$f_{HP}$, $f_{LP}$]. The default values are $f_{LP}=f_{Dmin}=f_{sampling}/5$ and $f_{HP} = f_{Dmax} = f_{sampling}/500$. Both $f_{LP}$ and $f_{HP}$ can be selected at will: their impact is examined in section 3.3.2.

In the case of a true Doppler burst, a clear maximum is found in the Fourier transform and that peak frequency is extracted. For the final analysis, one goes back to the temporal signal. First, the raw signal is band-pass filtered around that peak frequency with a Butterworth filter of order 2 those -3dB bandwidth is set at 0.2 and 1.8 times the peak frequency to remove both some noise at high frequencies and the baseline component at low frequencies. Such a filtered Doppler signal is exemplified Fig.16. On that filtered signal, successive fringes are identified using a local maxima search. An instantaneous period is computed for each fringe as the elapsed time between successive local maxima. The Doppler burst is validated only if one can find n consecutive periods those values are the same within ε%, with ε ranging from 5 to 20%. When it is so, the velocity $V_i$ of the $i^{th}$ bubble is computed from eq.(1) using the average period determined over the n oscillations: the selection of suitable values of n and ε is discussed in section 3.3.2. Let us underline that experiments in bubbly flows have shown that the Doppler frequency is often decreasing when a water/air interface is approaching, possibly because of liquid inertia and localized interface deformation. With usual optical fibers, the interface deformation at a water-to-air transition gas is about 7-11µm for velocities below 10m/s (Liju et al., 2001): such a magnitude is compatible with the present observations since the decrease in frequency is typically observed over a few fringes (Fig.6). To get rid of that, the search for the n successive periods starts between 0 to 5 fringes away from the probe-interface contact. From now on, all the results presented exploit the Doppler modulation detected at air-to-water transitions ($n_{ext}$ =1). Hence, the final information provided by the signal processing consists in the arrival time and the gas dwell time of all detected gas inclusions plus the velocity for a subset of these bubbles. This is the same information available collected from multimode fiber probes, and thus, the same post-processing can be applied to extract variables of interest including concentration, number density, fluxes and interfacial area density (Cartellier, 1999).

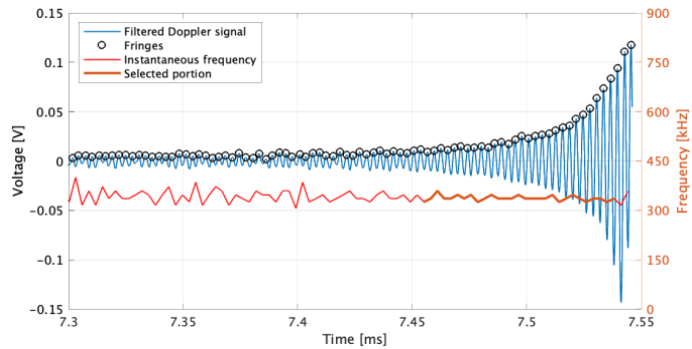

Fig.16: On a band-pass filtered Doppler burst (blue curve) observed before gas-liquid transition, fringes are identified (black dots) and assigned an instantaneous frequency value (red curve). The probe-interface contact is located at the extreme right of the signal: the search for consecutive fringes starts between 0 to 5 fringes away from that contact. The bold portion of the red curve corresponds to n=30 consecutive fringes with the same period within ε=8%: it is exploited to compute the average frequency and the bubble velocity.

To investigate the sensitivity of velocity measurements to the four processing parameters, namely n, ε and the filter bandwidth [$f_{HP}$, $f_{LP}$], we considered one superficial velocity in the homogeneous regime and another one in the heterogeneous regime. The main characteristics of the signals and of the processing are given Table 2: both records were long enough to ensure a satisfactory convergence. For that analysis, we considered the statistics built on detected velocities only without using any interpolation scheme to evaluate missing velocity data.

| Measuring conditions | Homogeneous conditions Vsg = 1.3 cm/s | Heterogeneous conditions Vsg = 16.25 cm/s |
|---|---|---|
| Record duration | 837 seconds | 95 seconds |
| Number of bubbles detected | 9315 | 9143 |
| Liquid level | 10 mV | 10 mV |
| Gas level | 0.86 Volts | 1.02 Volts |
| Peak to peak noise amplitude at liquid level | 10 mV | 15 mV |
| Peak to peak noise amplitude at gas level | 50 mV | 50 mV |
| Lower detection threshold | 50 mV | 50 mV |
| Upper detection threshold | 0.56 V | 0.68 V |
| Sampling frequency | 62.5 MHz | 62.5 MHz |
| High pass frequency $f_{HP}$ - Low pass frequency $f_{LP}$ (dynamic range of the filter $f_{LP}/f_{HP}$) | 140 kHz – 2137 kHz (dynamic range 15.3) | 140 kHz – 6250 kHz (dynamic range 44.6) |

Table 2: Conditions exploited to investigate the sensitivity of velocity measurements to processing parameters.

*3.3.1 Optimum choice of n and ε*

We examined first the mean bubble velocity and its standard deviation for a number n of periods ranging from 10 to 100 and for a maximum deviation between 2% and 25%. The reference values are set to n=20 and ε=10%: they correspond to standards in Laser Doppler Velocimetry. For the heterogeneous regime, the deviation (in %) from the reference is provided Table 3. It is color-coded: white data are within ±5% of the reference, grey data within ±10% and dark grey data correspond to deviations above ±10%. Clearly, there is always a significant range of processing parameters over which the measurements remain nearly the same. The mean velocity is recovered within ±5%, for ε between 10% and 20% and for n from 20 to 75. When considering the standard deviation, that interval slightly diminishes to ε between 10% and 15%. Similar results (not shown) were obtained in the homogeneous regime. Hence, high fidelity measurements are ensured provided that n and ε remain within the above-mentioned intervals.

The largest (above 15-20%) deviations systematically occur for extreme values of n and ε. That occurs when ε is too small (≤ 2%) resulting in a too strong selection of "stable" Doppler signals, or when ε is too large (≥ 25%) in which case the noise is easily interpreted as a Doppler signature. That also arises for n below 20: in that case, the search for short Doppler signals favors the validation of noise (see Fig.18 and the associated discussion). The worst cases occur for a small number of periods (n=10) and for ε in the upper portion of its range (20-25%): the deviation increases up to 30-60% for the mean velocity and it could exceed 100% for the standard deviation.

| Period homogeneity -> number of periods : | 25% | 20% | 15% | 10% | 5% | 2% |
|---|---|---|---|---|---|---|
| 10 | 60.03 | 28.64 | 6.33 | 0.67 | 2.17 | 16.15 |
| 20 | 29.09 | 4.46 | 0.88 | 0.00 | 3.02 | 20.82 |
| 30 | 14.41 | 1.51 | 0.07 | 0.64 | 5.63 | 23.57 |
| 40 | 7.99 | 0.57 | 1.73 | 4.09 | 6.30 | 22.10 |
| 50 | 1.82 | 1.42 | 2.91 | 3.39 | 6.21 | 22.48 |
| 75 | 0.19 | 2.48 | 3.81 | 4.19 | 6.69 | 28.72 |
| 100 | 0.07 | 3.97 | 5.46 | 5.97 | 5.81 | 29.85 |

| Period homogeneity -> number of periods : | 25% | 20% | 15% | 10% | 5% | 2% |
|---|---|---|---|---|---|---|
| 10 | 166.48 | 106.40 | 33.09 | 5.64 | 3.74 | 25.95 |
| 20 | 106.28 | 21.88 | 4.49 | 0.00 | 3.81 | 35.69 |
| 30 | 64.07 | 11.26 | 4.50 | 0.60 | 6.98 | 42.60 |
| 40 | 41.91 | 6.56 | 2.23 | 1.49 | 9.06 | 44.49 |
| 50 | 18.58 | 5.71 | 2.39 | 2.26 | 11.59 | 48.82 |
| 75 | 11.88 | 2.07 | 0.10 | 4.21 | 14.86 | 56.64 |
| 100 | 12.23 | 3.38 | 0.94 | 7.86 | 14.31 | 56.75 |

Table 3: Deviation in % from the reference value of the mean inclusion velocity (top) and of its standard deviation (bottom) to processing parameters (Heterogeneous regime Vsg=16.25 cm/s, signal characteristics as in Table 2).

The validation rate, defined as the fraction of detected bubbles for which information on velocity is directly available, is shown Fig.17 as a function of n and ε. It monotonously decreases with n, i.e. when increasing the minimum duration of Doppler bursts. Whatever n, it is almost unaffected by ε for ε between 10% and 20% but it drastically drops for ε below 5%: such trends are similar to those observed in LDV response (Durst et al., 1976). To maximize the validation rate while ensuring a good fidelity, selecting n=20 periods and ε=10% seems to be an optimal choice. For these values, the validation rate reaches 66% in homogeneous conditions, and it decreases down to about 50% in heterogeneous conditions, probably because of the complex bubble trajectories in that regime. Let us note that all the results presented in this section correspond to a probe pointing downward, that only detects bubbles moving upward. The detection of bubbles flowing downward will be examined in section 4.

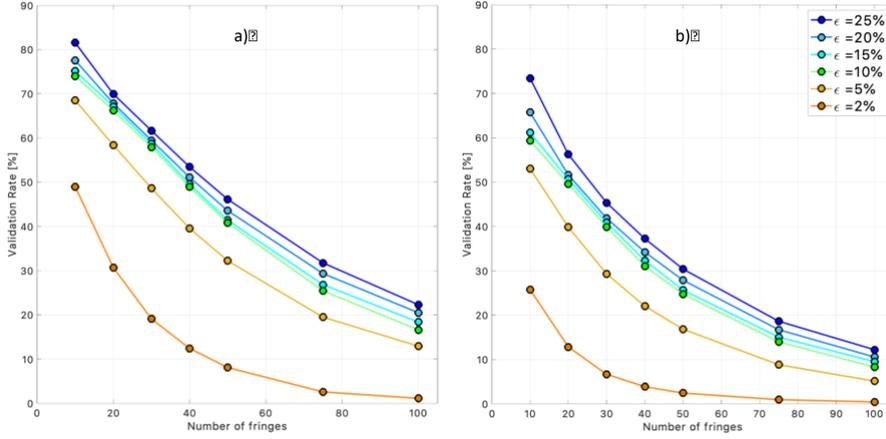

Fig.17: Velocity validation rate versus the number n of Doppler periods for various values of the maximum deviation ε for signals acquired on the column axis: a) in the homogeneous regime (Vsg=1.3cm/s) and b) in the heterogeneous regime (Vsg=16.25 cm/s) - signal characteristics as given in Table 2.

We also analyzed the impact of n and ε on velocity distributions. Examples are provided Fig.18 in the heterogeneous regime. In Fig.18-left, the maximum deviation ε is fixed to 10% and the number of successive periods evolves from 10 to 100. Since, for a given record duration, the number of detected velocities decreases with n (Fig.17), the distributions are smoother at small n. For the same reason, the dynamics of the pdfs increases when diminishing n: it reaches 2.5 decades for n=10. Aside from this, all velocity distributions for a given regime happen to be almost identical when varying the number of periods. In Fig.18-right, n=20 and the deviation evolves from 2% to 25%. Again, all distributions collapse reasonably well, except for the appearance of a finite probability to detect large velocities when ε exceeds 20%. These large velocities are within the interval 4-7m/s (see brown and red curves): they are well beyond the main set of detected velocities. When going back to the corresponding raw signals, it appears that the peak amplitude of their Fourier transform was systematically lower than for true Doppler bursts. Beside, a measure of velocity was successfully performed for ε large (≈25%) but not for ε small (≈10%): clearly, these series correspond to noise unduly validated as Doppler signals. They are also responsible for the huge deviations observed in Table 3 at small n and large ε. These incorrect measurements disappear from the statistics when ε is small enough, typically about 10%. The same behavior holds in the homogeneous regime. In conclusion, the recommended range of processing parameters corresponds to 10% ≤ ε ≤ 15% and n=20. This range ensures a low (≈5%) relative uncertainty on the mean velocity and on its standard deviation as well as a good reproducibility of velocity pdfs together with an optimized validation rate. Let us also underline that, with such criteria, the maximum acceleration $\gamma_{max}$ of the inclusion for which a velocity measurement is feasible is set by the maximum velocity difference over the signal duration, i.e. $\gamma_{max} = (\varepsilon/n) V^2 / (\lambda_0/2n_{ext})$. With the selected n and ε, that maximum acceleration $\gamma_{max}$ equals 10g for an inclusion travelling at V=0.1m/s and 100g for an inclusion travelling at V=1m/s. These magnitudes are significantly larger than the acceleration in the liquid deduced from Pavlov time series (the maximum liquid acceleration measured with Pavlov tubes was ±20m/s$^2$ in the heterogeneous regime), so that the velocity statistics gathered in bubble columns are not biased because of some limitation on the acceleration of inclusions.

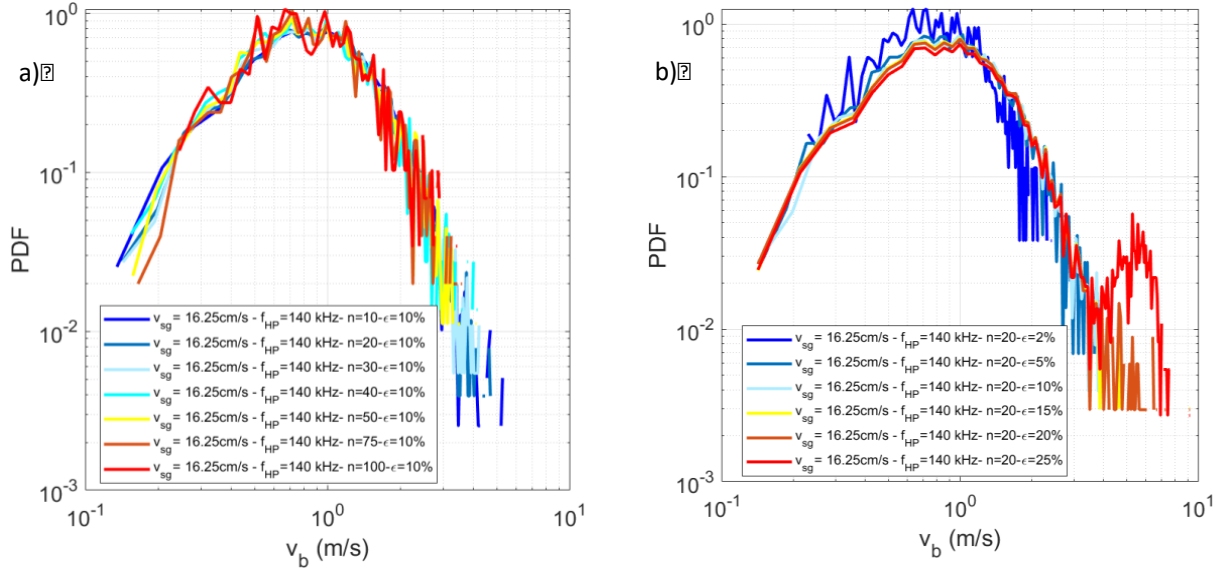

Fig.18: Evolution of bubble velocity distributions as a function of the parameters n and ε ☐☐ ☐☐e ☐eterogeneous regime (Vsg=16.25cm/s): a) impact of the number of periods n from 10 to 100 for a fixed deviation ε=10%, b) impact of the deviation ε from 2% to 25% for a fixed number of periods n=20 - Signal characteristics as given in Table 2 - $f_{HP}$=140 kHz.

*3.3.2 Optimum choice of filters*

The parameters n and ε being set, the settings of the band-pass filter [$f_{HP}$ ; $f_{LP}$] are worth to be examined, as we have seen that incorrect velocity data can be extracted from noise. Concerning the low pass frequency $f_{LP}$ that controls the upper limit in velocity, the maximum velocity is correctly detected when $f_{LP}$ is higher than about 0.8 times the highest Doppler frequency $f_{Dmax}$ in the record (this 0.8 coefficient is slightly less than unity because of the shape of the filter). Concerning the lower velocity limit, low velocities progressively disappear from the pdfs as $f_{HP}$ increases: the cut-off occurs at a frequency about 0.8-0.9 times $f_{HP}$, in agreement with the characteristics of the Butterworth filter. However, we also analyzed the impact of the actual velocity dynamic range on these limits using signals collected in the homogeneous and in the heterogeneous regimes. The following trends were sorted out from this investigation (see Mezui M'Obiang Y., 2021 for more details):

- The detection of the maximum velocity is rather straightforward. Whatever the filter dynamic with respect to the actual velocity, the maximum velocity is successfully determined provided that the low pass filter is such that $f_{LP}/f_{D\ max} \approx 1$ so that the highest velocities in the signal are not rejected. More precisely, $f_{LP}/f_{D\ max} \geq 0.8$ suffices for an accurate detection of the true maximum. Note also that $f_{LP}$ should not be increased too much as the detection of the maximum velocity deteriorates if the low pass frequency exceeds more than about three times the maximum Doppler frequency in the signal.
- The detection of the minimum velocity remains accurate, unless the low-pass frequency is too high compared with the maximum Doppler frequency, say for $f_{LP}/f_{Dmax} \geq 1.3$. Also, low velocities data start to be lost when, for a given low pass filter, the dynamic range of the band-pass filter becomes too small compared with the actual frequency dynamics in the signal. To correctly detect the minimum velocity when $f_{LP}/f_{Dmax} = 0.8$, the filter dynamic $f_{LP}/f_{HP}$ should be at least equal to the actual frequency dynamics in the signal.

When these recommendations are fulfilled, the validation rate happens to be maximized. We also checked that the mean velocity $V_G$ and its standard deviation $V_G$' are nearly unaffected by the filter settings when the latter are properly chosen. For example for a low-pass filter set at $f_{LP}/f_{D\ max}$=0.8 and for a filter dynamics $f_{LP}/f_{HP}$ between 1 and 10 times the actual signal dynamics, the variation is 0.25% on $V_G$ and 1.4% on $V_G$' in the homogeneous regime. In the heterogeneous regime, the variation is 4.9% on $V_G$ and 1.55% on $V_G$'. All these figures represent quite acceptable sensitivities.

Let us also underline that the discussion on n and ε parameters was achieved for a high pass frequency $f_{HP}$ set to 140kHz (Table 2). That choice was indeed within the recommendations made on the filters settings as the filter dynamics was 1.36 times the frequency dynamics in the signal for the homogeneous regime (and 1.1 times for the heterogeneous regime). Therefore, the recommendations made for n and ε correspond to true optimal values that are not filter dependent.

At this stage, it is interesting to comment on extreme values of the measured velocity. Over a record of 4500 valid velocity measurements in heterogeneous condition, two data were found at 7m/s (see Fig.19). If these data are set apart, the next largest velocity detected is about 5m/s, a value that corresponds to the visual limit of the pdf's tails shown Fig.19. We checked that the two data around 7m/s do correspond to good quality Doppler signals and that they are not artifacts. It seems odd to detect velocities as large as almost 10 times the mean. It is likely that such events come from some localized interfacial deformation that may occur because, for example, of bubble shape oscillation or during an isolated and rare coalescence event. In such cases, the measured velocity is probably set by the geometrical interface displacement velocity da/dt as discussed in section 2.2.2: the latter can indeed be quite large when the interface dynamics is controlled by inertia and surface tension. In the homogeneous regime, rare events at high velocities were also observed, but these velocities remained moderate (about twice the mean) and they were probably corresponding to a mean bubble motion instead of an interface deformation.

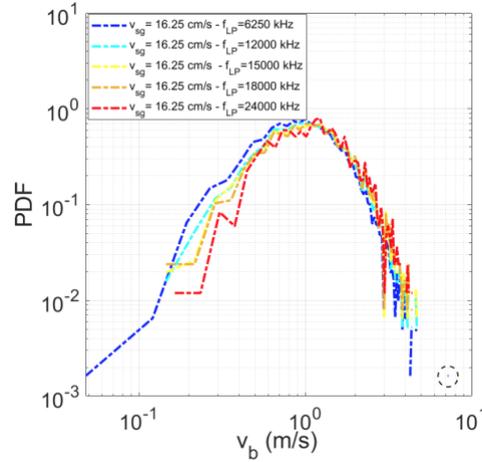

Fig.19: Evolution of the detected velocity pdfs in the heterogeneous regime (Vsg=16.25cm/s) with the low pass frequency $f_{LP}$ of the band-pass filter, all others parameters being kept constant ($f_{HP}$=30kHz, n=20, ε=10%, signal characteristics as in Table 2). The largest velocity recorded is circled in the figure.

Overall, the above analysis demonstrates that the signal processing is robust and that it successfully handles a velocity dynamic of half a decade in a single processing step. Note that the maximum dynamic range that the proposed routine can handle in a single step has not been identified. In case of a frequency dynamics exceeding the system limit, one could implement a multi-pass processing using band-pass filters shifted to lower frequencies in order to accurately detect the smallest velocities. In practice, the signal characteristics are usually not known a priori. It is thus recommended first to adjust the low-pass filter to properly detect the largest Doppler frequencies present in the flow, and then to adjust the high pass filter until the filter dynamic range matches that of the signal. With such a processing, the resolution on Doppler period measurements is about $f_D/(n\ f_{sampling})$: as n is set to 20, and $f_{sampling}$ is at least 5 times $f_{D\ max}$, the resolution on velocity is about 1/100 of the largest velocity.

Finally, it is worth commenting on the system performances in terms of the minimum detectable velocity. Indeed, as in LDV, the Doppler frequency is linearly related with the measured velocity. Therefore as the velocity approaches zero, the Doppler frequency becomes harder to be measured since the Doppler period increases without bound. This is why LDV systems usually incorporate a frequency shift of the light wave (with Bragg or Pockels cells) so that the zero of the Doppler frequency is shifted by a fixed amount. In these circumstances, positive, negative as well as zero velocities in the laboratory frame of reference can be identified. Since the light frequency is not shifted in the Doppler probe, this sensor is unable to distinguish between positive and negative velocities. It is also unable to measure velocities too close from zero. Yet, there is no absolute limit on the minimum velocity detectable as far as the Doppler signal to noise ratio remains large and that Doppler signals are long enough. For example, 1cm/s velocities were detected on localized interfaces (Fig.5), and bubble velocities down to 3.4 cm/s were recorded in the heterogeneous regime at Vsg=16.25cm/s.

3.4 *Chords measurements*

The processing extracts the gas residence time $T_G$ for all signatures. For some of these, it also provides the bubble velocity $V_b$. When these two information are available for the same event i, one can infer the gas chord $C_i = V_{bi} T_{Gi}$ detected by the probe. The set of these chords provides the so-called raw chord distribution as shown Fig.20. We analyzed the sensitivity of the raw chord distribution to the processing parameters n and ε. As for velocity, there is always a significant range of n and ε over which the mean chord and its standard deviation remain nearly the same. More precisely, and whatever the flow regime, the mean chord is the same within ≈5% for ε between 5% and 15% and for n from 10 to 50. The same conclusion holds for the chord standard deviation when considering a 10% variation. These figures are pretty much the same as for velocity. This is partly expected because the distributions of $T_G$ do not depend on n, nor on ε. Yet, the fact that the sensitivities to n and ε are similar for chords and for velocities means that there is no significant correlation between velocity and gas dwell time measurements: in other words, no bias in gas velocity statistics is to be expected from the Doppler probe.

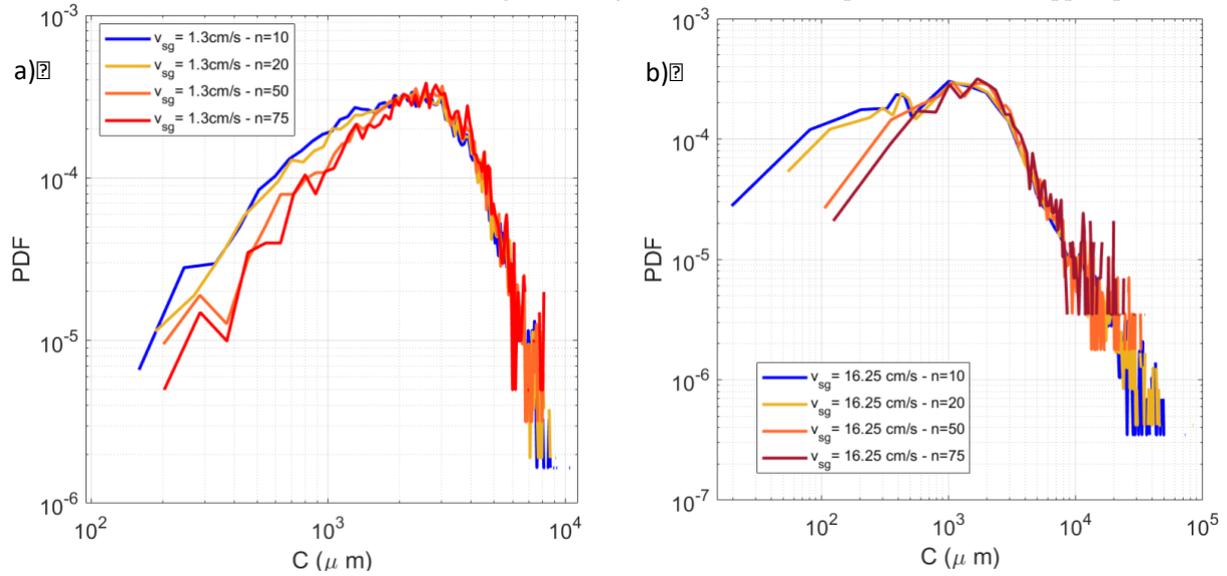

Fig.20: Evolution of raw chord distributions (i.e. without velocity interpolation) with n for a fixed ε=10%: a) in the homogeneous regime (Vsg=1.3cm/s) and b) in the heterogeneous regime (Vsg=16.25cm/s). Signal characteristics as given in Table 2 with $f_{LP}$=92kHz, $f_{HP}$=1.005MHz for the homogeneous condition, and with $f_{LP}$=146kHz, $f_{HP}$=3.751MHz for the heterogeneous condition.

As shown Figure 20-right, the minimum chord directly detected (i.e. with direct velocity measurements) increases with the number n of periods. This is indeed expected as a valid velocity measurement requires a minimum number n of successive periods, that is a minimum signal duration within the gas phase: that duration corresponds to a length $L = n \lambda_0/(2n_{ext})$. L is about 8μm for the smallest n considered (n=10): this is larger than the latency length $L_S$, so that the corresponding signals exhibit a flat plateau at the gas level and a neat Doppler modulation. Consequently, L also represents the shortest chord that can be detected with direct velocity detection.

The only puzzling result we observed concerns the maximum chord detected in the heterogeneous regime. The latter nearly reaches 10cm at Vsg=16.25cm/s, to be compared with a mean chord about 4mm and a standard deviation about 7mm. That result was insensitive to n, ε and to filter limits. We scrutinized the flow inside the column with an endoscope. The key point is that no bubbles with such sizes were present in the column: the largest bubble ever observed was 9mm in size. The huge chords detected with the probe represent a few percent of the total population. A close examination of the associated signals indicates that more than 60% of these unambiguously show multiple bubbles, as illustrated Figure 21. If one considers successive bubbles separated by a liquid film that is thinner than the probe latency length, then, the signal will never reach the liquid level. Consequently, a single bubble would be detected with a gas residence time equal to the sum of all constitutive gas dwell times. In Figure 21, that process concerns 3 bubbles in signal a) and 6 (and possibly more) bubbles in the signal b). It is therefore highly likely that the huge chords detected in heterogeneous conditions do arise because of close-by bubbles that are not individually distinguished by the probe. The companion question is where do these close-by bubbles come from? It is likely that such events correspond to bubbles trapped in the core of very intense vortices that form in the strongly sheared regions of the flow, the latter being the consequence of unsteady large-scale motions arising in the heterogeneous regime (Maximiano Raimundo et al., 2019). Indeed, zones at high void fraction, up to ten times or more the mean gas hold-up, do exist in the heterogeneous regime (Maximiano Raimundo et al., 2019) and that may explain how multiple bubbles can be artificially associated to provide apparent very long chords. The detection of such large chords remains rare enough, so that they do not alter the mean chord but they significantly contribute to the standard deviation. To quantify this effect, we considered the contribution of chords larger than the mean value $C_G$ plus two standard deviations $C_G'$. In the heterogeneous regime at Vsg=16.25cm/s, $C_G$=4mm and $C_G'$=7.6mm. The chords above $C_G+2C_G'$=19mm represent 3% of the population: their contribution to the mean value is 3% while their contribution to the standard deviation amounts to 20%. In conclusion, and as for velocity, the recommended processing parameters ensure a reasonable uncertainty on the mean chord and on its standard deviation as well as a good reproducibility of raw chord distributions.

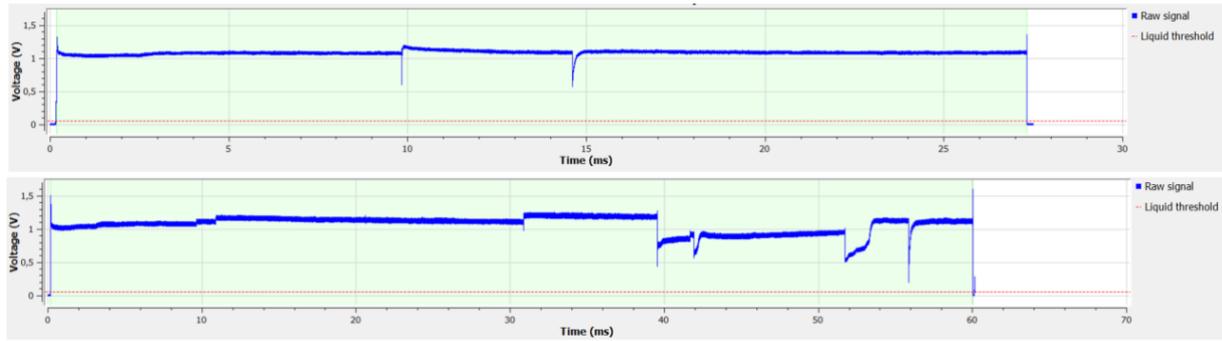

Fig.21: Signals corresponding to long gas chords collected in the heterogeneous regime: top) chord = 43mm at a velocity ≈ 1.6 m/s; bottom) chord = 84mm at a velocity ≈ 1.4 m/s.

**4. First exploitation of the Doppler probe in a bubble column operated in the heterogeneous regime**

To test the capabilities of the Doppler probe, we exploited that sensor in an air/water bubble column operated in the heterogeneous regime in order to gather some new information. The experiments were performed in a 3m high bubble column with an internal diameter D=0.4m. The static liquid height was fixed to 2.02m that is about 5D, and the measurements were collected in the quasi fully developed region of the flow (Maximiano Raimundo et al., 2019). Beside, all the results discussed in section 3 were gathered with a vertical and downward oriented probe. However, it is known that heterogeneous conditions are associated with large velocity fluctuations. In particular, and especially off axis, the liquid can experience downward directed motion over significant periods of time. As the Doppler sensor is not responding to negative velocities, it is thus essential to acquire information from an upward oriented probe and to combine it with the information gathered from a downward oriented probe. Xue et al., 2008 used a similar procedure with four-tip sensors. Let us examine its relevance for Doppler probes.

4.1 *Sensitivity to the main flow orientation and measurements of bubble velocity pdfs*
To complement the tests presented in section 2.2.3, let us first analyze the sensitivity of velocity measurements to probe orientation in the complex bubbly flows arising in a bubble column. These measurements were collected at H/D=3.62 above injection, where the flow is quasi fully developed. In these experiments, the optical fiber was inserted in a 0.9mm internal diameter and 1.5mm outer diameter stainless steel tube, and the fiber protruded 7 mm out of the tube. To avoid vibrations of the sensor, this tube was held on a 9° wedge support (10mm thick, lateral extension 60mm long), itself fixed at the extremity of a square (width 20mm) metallic bar immersed in the column from the top. The tube was located at the extremity of the wedged support (55mm away from the vertical bar), and the optical tip of the probe was 76mm ahead of the wedge. Signals were collected with the same probe downward and then upward oriented. They were treated using optimum filters and standard values for n and ε. The results are provided Table 4.

| Flow conditions | Homogeneous (Vsg= 1.3 cm/s) | | Heterogeneous (Vsg=16.25 cm/s) | | | |
|---|---|---|---|---|---|---|
| Radial position | On axis | | On axis | | 0.7 R | |
| Probe orientation | downward directed probe | upward directed probe | downward directed probe | upward directed probe | downward directed probe | upward directed probe |
| Number of inclusions detected | 9297 | 6007 | 9143 | 6766 | 10505 | 3708 |
| Measuring duration | 837 s | 850 s | 95 s | 95 s | 221.8 s | 95 s |
| Detection frequency Hz | 11.1 | 7.1 | 96 Hz | 71 Hz | 47 Hz | 39 Hz |
| Velocity validation rate | 66.5% | 0% | 52.1% | 1.27% | 26.7% | 17.8% |
| Mean velocity m/s | 0.44 | - | 1.2 | - 0.40 | 0.86 | - 0.51 |
| Mean chord mm | 2.7 | - | 4.10 | 1.86 | 3.67 | 2.34 |
| Volumetric flux (m/s) | 0.026 | - | 0.208 | 0.0027 | 0.0548 | 0.0177 |

Table 4: Statistics gathered from the same probe using downward and upward orientations at H/D=3.625. The values presented correspond to a filter dynamic that is twice the signal dynamic defined at the 1‰ limits.

In the *homogeneous regime* (Vsg=1.3cm/s), all bubbles are travelling upward in a laboratory frame. They thus produce head-on impacts on a downward oriented probe. Consequently, the velocity validation rate is high: here, it equals 66.5%. When the same probe is oriented upward, two effects are noticeable:
- First, there is a decrease in the detection frequency compared with a downward directed probe. This is probably due to the rather thick probe holder used here that alters some bubble trajectories and thus hampers the correct detection of all incoming bubbles. Here, an upward directed probe detects 64% of incoming bubbles: that is the decrease in the detection capability due to an inverted probe orientation amounts to 36%. The decrease in terms of void fraction is of the same order, about 40%.

- Second, when the probe is upward oriented, the velocity validation rate drastically drops. In the example of Table 4, no Doppler signal was collected among more than 6000 bubbles detected. In another record, the interface velocity was measured on a single event out of 2563: that corresponds to a 0.04% success rate. These results confirm the conclusion of Section 2, namely that the Doppler probe is unable to detect negative velocities with respect to the probe orientation.

In the *heterogeneous regime*, and depending on Vsg and on the radial location, the liquid and/or the bubbles can experience both up and down motion in a frame attached to the column (Xue et al., 2008; Maximiano Raimundo et al., 2016; Guan et al., 2017). For that reason, the response of the probe was analyzed at two locations, on the axis of the bubble column and at a lateral position x/R=0.7 where the mean liquid velocity is close to zero. The investigation was done at Vsg =16.25cm/s.

- On the column axis, the detection frequency decreases from 96Hz to 71Hz when the probe is inverted with respect to the mean flow. According to the liquid velocity pdf collected from a Pavlov tube shown Figure 22, the liquid flow on the axis is upward directed 96% of the time. Therefore, and as in homogeneous conditions, the decrease in the detection frequency is most probably due to the probe holder that hampers the detection of some of the incoming bubbles. For the holder used and the flow conditions considered in Table 4, the decrease in the detection frequency when using an inverted probe orientation amounts to 26%. In terms of void fraction, the decrease is 10 to 20%: these figures are slightly less than in the homogeneous regime.

- We have just seen that negative liquid velocities are recorded 4% of the time on the axis, and that their magnitude is up to 0.5m/s (Fig.22). Such velocities are strong enough to entrain some bubbles downward. When the Doppler probe is upward directed, it indeed detects a few meaningful Doppler signals that amount for 1.27% of the number of detected bubbles (Table 4): these signals represent 86 events out of 6766 detected bubbles within 95 seconds. When the probe is downward directed, the velocity validation rate is close to 50% leading to more than 4650 valid velocity data collected within the same measuring duration of 95 seconds. Hence, the Doppler sensor detects a downward flow 86 times over a set of 4650 events: that represents about 1.8% of the time. This is consistent with the statistics on the liquid phase provided by the Pavlov tube: indeed, that figure closely matches the time fraction (≈2%) corresponding to a liquid velocity that is directed downward and that exceeds the terminal bubble velocity (the latter has been set to 0.25m/s here).

- Let us now consider the measurements performed at the lateral position 0.7 R. The liquid has a mean velocity close to zero. As indicated by the liquid velocity distributions shown Fig.22, it is upward directed roughly half of the time and downward directed roughly half of the time. That feature explains why in Table 4, the detection frequencies are nearly the same for the two probe orientations. That feature also explains why these detection frequencies (respectively the velocity validation rates) are about one half the detection frequency (respectively the velocity validation rate) corresponding to a downward directed probe on the column axis where the flow is 96% of the time (that is nearly always) directed upward.

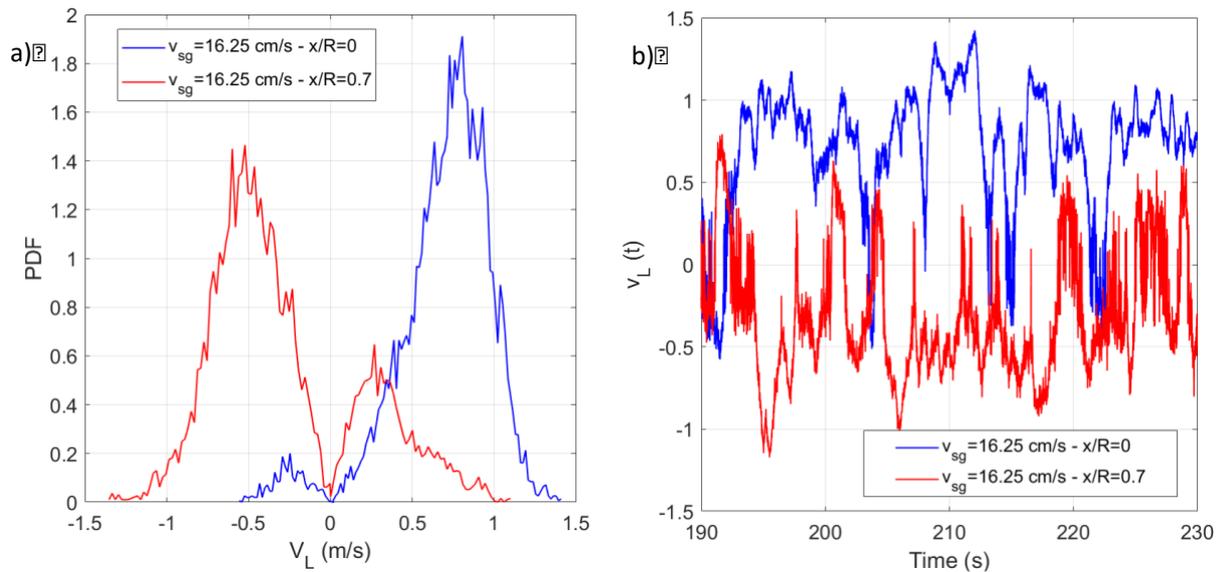

Fig.22: Unconditional liquid velocity distributions in the center of the column and at a radial position 0.7R measured by a Pavlov tube (a), and time series recorded from Pavlov tube (b). The Pavlov tube consisted of long tubes (6mm external diameter, thickness 0.5mm) plunging in the column from the top. Their extremity was bent as an "L". The horizontal portion of the "L" was 150mm long. Each tube was pierced with a 0.5mm diameter orifice located 5.5cm from the extremity of the "L". The two orifices were located along a vertical, and the distance between them was 12mm. The sensor was a Rosemount reference 2051 CD2 able to measure velocities in the range ±3.1m/s with a resolution of 0.14m/s. Data collected at H/D=3.62 for Vsg=16.25 cm/s.

Taking advantage of the above observations, the question to address now is how to reconstruct the bubble velocity distribution at a given location in the column from the information collected from the two orientations of the same probe. The most straightforward procedure consists in assembling the data gathered for each orientation using the same measuring duration. However, two conditions must be fulfilled for this procedure to be valid: i) the two sets relative to the two probe orientations should be independent from each other, and ii) for each set, the statistics should not be biased. These two conditions are scarcely analyzed in the literature, and it is worth to discuss them for the Doppler sensor.

- Concerning the first condition, we have shown that the Doppler probe only detects positive velocities with respect to the probe orientation. Hence, there is no mixing between positive and negative velocity sets as each set corresponds to a different probe orientation: these two sets are truly statistically independent and one can indeed sum up the information collected for each probe orientation.

- The second condition is also fulfilled because, as shown in section 3, the proposed signal processing is reliable for a given probe orientation, and also because the analysis of Table 4 does not evidence any bias related with a change in the probe orientation.

Gas velocity distributions can be therefore built by aggregating direct velocity measurements collected from the two probe orientations using the same measuring duration. In the following, these are named "combined direct gas velocity distributions". Examples of such distributions are shown Figure 23 for data collected on the axis and at a radial position 0.7R.

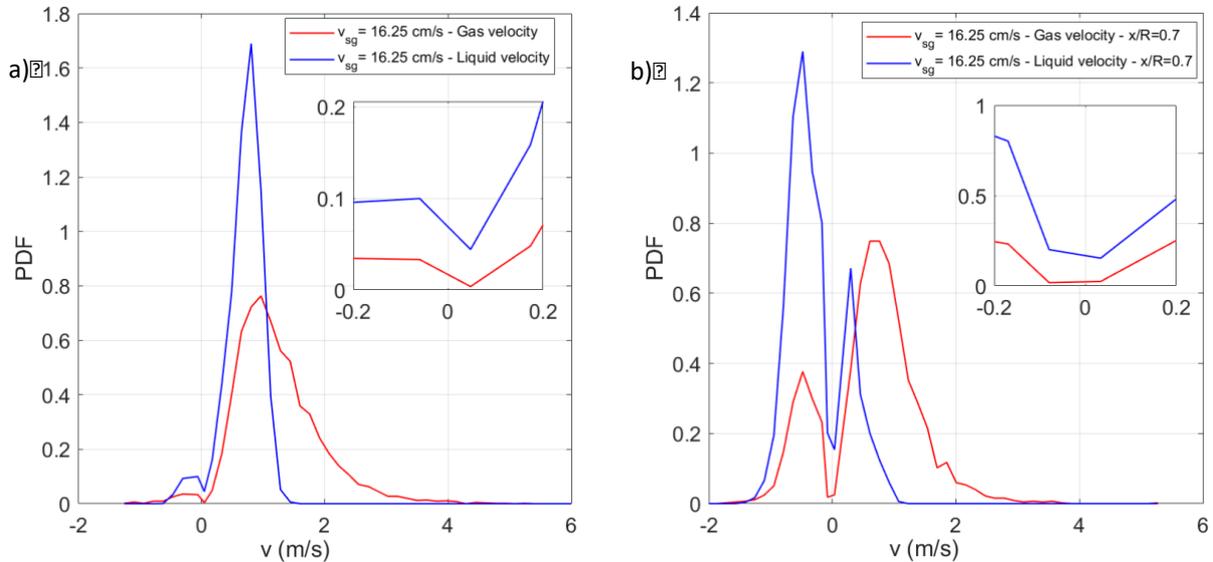

Fig.23: Combined direct gas velocity distributions at radial positions 0 (a) and 0.7R (b). Liquid velocity distributions measured with the Pavlov tube at the same locations are also shown. Measurements at H/D=3.625 for Vsg=16.25 cm/s. The inserts are magnification of the curves in the vicinity of the zero velocity.

Before pursuing our investigation of bubble velocity measurements, a comment deserves to be made on the distributions shown Fig.23. Indeed, a striking feature is the difference between the maximum velocities of the bubbles and of the liquid. Considering a probability density of $10^{-2}$, the velocity of bubble reaches 3.9m/s on the axis while the liquid reaches about 1.4 m/s. At x=0.7 R, these values are respectively 2.9m/s for bubbles and 1m/s for the liquid. Such large differences are unexpected: due to the continuity of normal velocities at interfaces, one expects instead to detect some liquid velocity events at the maximum velocity of bubbles. In our opinion, the fact that such events are absent from Fig.23 indicates that the Pavlov tube is not able to detect such events. This limitation is not due to the dynamic (±3.1m/s), nor to the resolution on velocity (0.14m/s) of the pressure sensor selected. The probable causes are two fold. On one hand, the temporal resolution is limited, as the pressure sensor has a bandwidth of 100Hz while the passage of a bubble corresponds to shorter time scales, of the order of a few milliseconds according to the measured bubble sizes and velocities. On another hand, due to the distance between the two holes of the Pavlov tube (the latter is 12 mm), the measured fluctuations of the liquid velocity are not truly local but they are spatially filtered. Both limitations lead to the disappearance of "too" fast or "too" local fluctuations in the measured distributions of the liquid velocity. In the perspective of improving the characterization of turbulence in bubble columns, it would be interesting to optimize Pavlov tubes design to better capture liquid velocity fluctuations, or to exploit alternative techniques less prone to temporal or spatial filtering of the liquid velocity field.

Going back to gas velocity measurements, the procedure proposed for reconstructing the gas velocity pdf is also supported by the following argument. All gas velocity distributions shown Fig.23 exhibit an interval around zero in which no bubble velocities are detected. Such a feature arises because the liquid flow never keeps a zero velocity for a sufficient long time. Indeed, the shifts between positive and negative liquid velocities are always quite fast as shown by the time series in Fig.22: the records from the Pavlov tube indicate a time scale below $10^{-2}$ seconds for the duration of these transitions. Millimeter size bubbles cannot adapt to such steep changes as the bubble response time is of the order of $a^2/\nu_L \approx 1$ second where a is the bubble radius and $\nu_L$ is the kinematic viscosity of water in ambient conditions. Note that this 1second time scale is underestimated as it has been evaluated here for a 1mm bubble radius while the equivalent bubble diameter is usually larger (it equals 7mm±0.3mm in our experiments). This is why bubbles never (or at least extremely scarcely) stick to a zero velocity or to a velocity too close to zero. To quantify that closeness, let us underline that the filters used in the signal processing have been systematically adjusted in order to detect the minimum velocities present in the signal: the latter happen to be ±5cm/s on the axis, and +3cm/s; -8cm/s at the radial position 0.7R. No bubble velocity was detected within these intervals around zero, and that observation holds both on the column center and at 0.7R. These "holes" observed around zero in the velocity distributions are thus meaningful and do not result from some drawback of the measuring technique. In addition, as the gas inclusion velocity is scarcely in the vicinity of zero, there is no ambiguity on the velocity direction of bubbles, and thus there is no mismatch between the two velocity sets resulting from the two probe orientations. In conclusion, the information gathered from a Doppler probe corresponds indeed to bubbles with a positive velocity (positive relative to the probe), and that statement applies to each probe orientation.

To close that section, let us mention that we attempted (but without success) to provide a velocity to all detected bubbles using a time interpolation routine as done previously (e.g. Cartellier, 1998). As the flow direction shifts from up to down in the heterogeneous regime, the difficulty is indeed to distinguish in a given record between positive and negative bubble velocities with respect to the probe. A probe of a given orientation detects an ensemble of successive bubbles with a positive velocity (with respect to the probe orientation) followed by an ensemble of successive bubbles with a negative velocity (with respect to the probe orientation) and so on. Bubbles with a positive velocity are detected, for a fraction of them their velocity is measured and, within that ensemble, a time interpolation routine can be implemented to evaluate missing velocities. On the other side, bubbles with a negative velocity can be detected but no velocity information can be extracted from their signatures, so that none of the bubbles with a negative velocity can receive a velocity (neither directly, nor by way of interpolation). The issue here is to distinguish between positive and negative velocities in a given record. As the Doppler sensor in its current configuration is unable to detect the sign of the velocity, we attempted an empirical procedure to identify ensembles of successive bubbles having the same velocity direction by examining the statistics on the waiting time between successive direct velocity detections. Yet, that criterion happens to be insufficient to develop a reliable routine. More precisely, a criterion on the waiting time leads to reasonable estimates of the mean and of the standard deviation of bubble velocity for positive velocities and for negative velocities considered separately. However, the relative weight of positive and negative velocities in the distribution remains quite sensitive to a threshold on the waiting time. Therefore, the combination of positive and negative data sets cannot be achieved on an undisputable basis. As a consequence, all the results presented in the following sections are based on combined direct velocity measurements.

4.2 *Comparison with bubble velocity pdfs available in the literature*

As discussed in the introduction, bubble velocity measurements in heterogeneous conditions are rare. Moreover, bubble velocity distributions are scarcely available. Comparisons are achieved here below with the data gathered by Xue et al., 2008 with a four-tip probe, and with those collected by Guan et al., 2017 with an horizontal bi-probe. They both used tap water and air under ambient conditions. For the comparison, we consider flow conditions as close as possible between theirs and ours, but these flow conditions are never the same. In addition, it is not clear how Xue et al., 2008 and Guan et al., 2017 processed their signals and how the statistics from the two probe orientations were combined. Consequently, it is difficult to evaluate the nature of the velocity that was measured (bubble, gas, arithmetic mean, phase average…), nor the potential bias. Thus, we will simply consider their data as they are, and we will compare them with our combined direct gas velocity distributions.

For Xue et al., 2008 we consider the data they acquired at a superficial velocity of 14cm/s in their D=0.162m column equipped with their sparger n°2. The data were collected at H=5.1 D above injection, i.e. within the quasi-fully developed region of the flow. Concerning bubble sizes, Xue et al., 2008 report a mean bubble chord (measured with their four-tip probe) from 5 to 6mm: that corresponds to a Sauter mean diameter along a vertical direction about 7.5 to 9 mm. These figures are larger than (but comparable to) the 6.2mm (respectively 7.6mm) Sauter mean diameter along a vertical (respectively horizontal) we measured at Vsg=16.25cm/s. In addition, their chord distribution indicates that bubbles up to 15mm were present in the flow. Hence, and contrary to our experiments, significant coalescence occurred in the conditions investigated by Xue at al., 2008. With these differences in mind, let us compare the bubble velocity pdfs in Figure 24. The original data of Xue et al., 2008 are somewhat scattered, so we represent here their mean trend as the average between the upper and lower envelopes of their data. On the axis, two superficial velocities close to the 14cm/s value of Xue et al., 2008, namely 13cm/s and 16.25cm/s, are considered to ease the comparison. Both techniques provide bimodal distributions that are "globally speaking" similar except that the pdfs extend up to much larger positive velocities with the Doppler technique than with the four-tips probe. Off axis, velocity pdfs are also globally similar even though we compare the data from the Doppler probe collected at x/R=0.7 for Vsg=13cm/s with the data from the four-tips probe collected at x/R=0.9 for Vsg=8cm/s. As on the column axis, the Doppler probe detects larger positive velocities than the four-tips probe. Beside, the two peaks of the distribution observed for positive and for negative velocities are almost symmetrical in Xue et al., 2008 while it is not so with the Doppler probe: this is possibly partly due to the difference in the lateral position as the proportion of up and down flows changes with the distance to the wall.

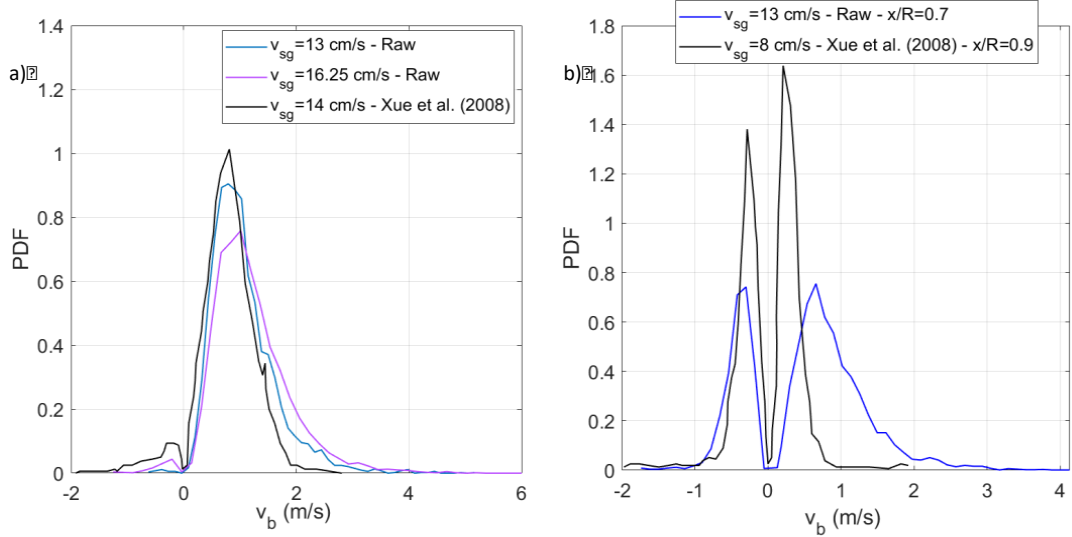

Fig.24 : Comparison of gas velocity distributions collected with a four-tip probe by Xue et al., 2008 (conditions: air-tap water, sparger n°2, column diameter D=16.2cm, H/D=5.1, Vsg=14cm/s) with the combined direct bubble velocity pdfs collected with a Doppler probe in the present experiments (air-tap water, column diameter D=0.4m, H/D=3.625, Vsg=16.25cm/s): a) on the column axis, b) off axis.

For the comparison with Guan at al., 2017, we considered the data they collected in their 15cm diameter column for a superficial velocity of 23cm/s and at H/D=5.3 that is within the quasi-fully developed region of the flow. The volume equivalent diameter of their bubbles was comprised between 4.5 and 6mm. There was no coalescence, or at least it was very weak. As shown Figure 25, the agreement with the data collected with the Doppler probe happens to be somewhat satisfactory. In particular, and contrary to the data from Xue et al., 2008, the extents of the peaks corresponding to positive velocities are similar at both radial positions. However, the negative velocities recorded by Guan et al., 2017 are much stronger that those measured with the Doppler probe. Moreover, as with Xue et al., 2008, there are noticeable differences on the heights of the peaks. These differences are possibly related to the way raw data delivered from the probes are post-treated - but very few details are provided by Xue et al., 2008 and by Guan et al., 2017 in that respect, and/or with potential bias of the four-tips probe or of the bi-probe techniques about which, as said in the introduction, little is known. Because of this lack of knowledge, it is difficult to pursue further the analysis, and in particular to quantitatively compare the probabilities to observe positive or negative (in the laboratory frame) velocities as a function of Vsg and of the radial position. In particular, we examined the first moments resulting from the data of Xue et al., 2008 in Table 5 and of Guan et al., 2017 in Table 6. Due to their bimodal character, three sets of data were considered for the analysis, namely positive velocities only, negative velocities only and the entire distribution. It happens that no agreement between the three techniques is found either on the mean or on the standard deviation, except on the relative turbulence intensity $V'_G/V_G$ when considering positive velocities on the column axis only.

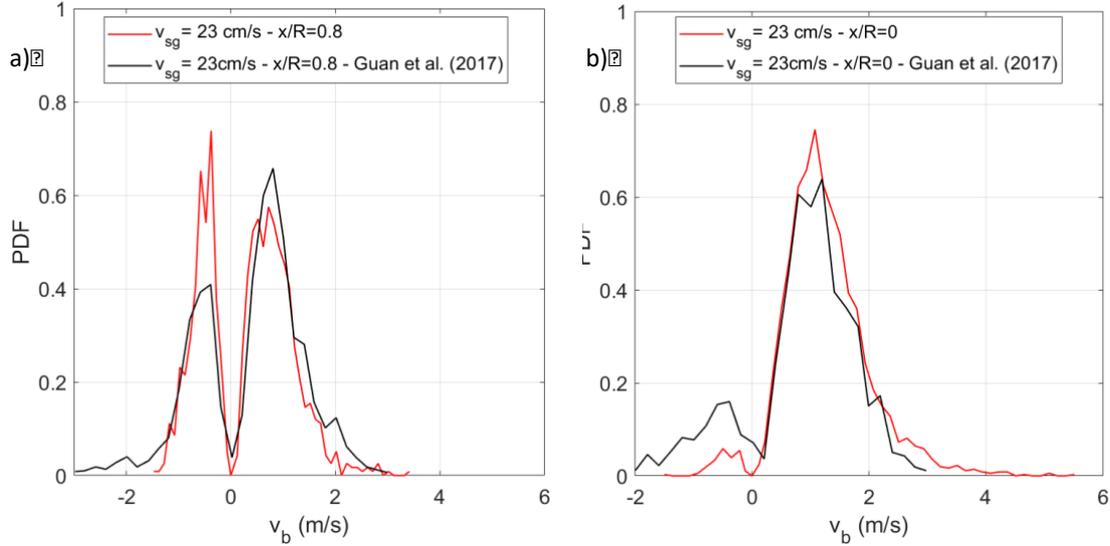

Fig.25: Comparison of gas velocity distributions gathered with a bi-probe by Guan et al., 2017 (air-tap water, column diameter D=15cm, H/D=5.3, Vsg=23cm/s) with combined direct bubble velocity pdfs collected with a Doppler probe in the present experiments (air-tap water, column diameter D=0.4m, H/D=3.625, Vsg=23cm/s): a) on the column axis, b) off axis.

|  | Mean velocity (m/s) | | Standard deviation (m/s) | |
| --- | --- | --- | --- | --- |
| On axis | Xue et al., 2008 | present data | Xue et al., 2008 | present data |
| Positive velocities only | 0.77 | 1.20 | 0.38 | 0.65 |
| Negative velocities only | -0.032 | -0.39 | 0.32 | 0.15 |
| All velocities | 0.73 | 1.19 | 0.51 | 0.64 |
| Off axis | Xue et al., 2008 (x/R=0.9) | present data (x/R=0.7) | Xue et al., 2008 (x/R=0.9) | present data (x/R=0.7) |
| Positive velocities only | 0.18 | 0.97 | 0.21 | 0.57 |
| Negative velocities only | -0.15 | -0.46 | 0.32 | 0.19 |
| All velocities | 0.023 | -0.07 | 0.43 | 0.56 |

Table 5: Various moments of the gas velocity distributions from Fig.24 measured by Xue et al., 2008 with a four-tip probe and measured with a Doppler probe. Experimental conditions are not strictly identical (see text).

|  | Mean velocity (m/s) | | Standard deviation (m/s) | |
| --- | --- | --- | --- | --- |
| On axis | Guan et al., 2017 | present data | Guan et al., 2017 | present data |
| Positive velocities only | 0.98 | 1.37 | 0.52 | 0.71 |
| Negative velocities only | -0.16 | -0.48 | 0.78 | 0.28 |
| All velocities | 0.81 | 1.32 | 0.99 | 0.77 |
| Off axis (x/R=0.8) | Guan et al., 2017 | present data | Guan et al., 2017 | present data |
| Positive velocities only | 0.52 | 0.95 | 0.46 | 0.53 |
| Negative velocities only | -0.22 | -0.56 | 0.57 | 0.27 |
| All velocities | 0.4 | 0.6 | 1.0 | 0.80 |

Table 6: Various moments of the gas velocity distributions from Fig.25 measured by Guan et al., 2017 with a bi-probe and measured with a Doppler probe. Experimental conditions are not strictly identical (see text).

4.3 *Unconditional statistics of the gas phase*

Owing to the encouraging performances of the Doppler probe, that sensor was used to explore the gas phase dynamics in the 0.4 m in diameter bubble column. We focus hereafter on unconditional measurements performed in the homogeneous regime as well as in the heterogeneous regime. Care was taken to ensure a satisfactory convergence of the statistics. Typically, for a probe facing downward, and depending on Vsg, from 5000 to 10000 bubbles were detected over measuring durations from 96 (at high Vsg) to 946" seconds (at low Vsg). To test the quality of injection, profiles were collected over a column diameter (from x/R=-0.9 to x/R=0.9) for Vsg =1.3cm/s: all variables including void fraction, bubble velocity (mean and standard deviation) and bubble size (mean and standard deviation) confirmed that the flow at Vsg =1.3cm/s was axisymmetric.

4.3.1 *Void fraction*

The local void fraction on the column axis $\varepsilon_G(0)$ was measured with a Doppler probe pointing downward. Indeed, and contrary to velocity data, the phase detection is only weakly sensitive to the probe orientation. In addition, the proper combination of the data collected from upward and downward probe orientations would require to properly weight bubbles moving up and down. That weighting process would require the knowledge of the phase detection error according to the orientation of the bubble velocity, an information we have not. This is why only a downward oriented probe was considered for void fraction measurements.

The void fraction on the column axis $\varepsilon_G(0)$ happens to linearly increase with the height above injector for H/D between about 1.5 up to 6, as shown Fig.26. Over that range of heights, the flow can be considered as quasi fully-developed. Moreover, the slope $d\varepsilon_G/dH$ monotonously increases with Vsg (Fig. 26-right). Assuming an isothermal system, with a constant bubble number density (no change due to coalescence and break-up), the axial change in the void fraction induced by the bubble expansion due to the axial pressure gradient writes $d\varepsilon_G/dH_{hydrostatic} = \varepsilon_G (1-\varepsilon_G) \rho_L g/P(H)$ where P(H) is the static pressure level at the measuring position. Assuming a hydrostatic pressure gradient, $d\varepsilon_G/dH_{hydrostatic}$ coincides with measurements of $d\varepsilon_G/dH$ in the homogeneous regime. On the opposite, in the heterogeneous regime, $d\varepsilon_G/dH_{hydrostatic}$ represents only 70% to 50% of the actual void fraction gradient. That feature indicates that another mechanism is at play that controls the axial increase of the void fraction. A tentative explanation could be that, among the bubbles that recirculate along the walls, some of them are captured by the upward flow all along the column height. In other words, the upward two-phase flow in the central region of the bubble column is continuously fed with a lateral gas flow rate all along the column height and not just at the bottom the column.

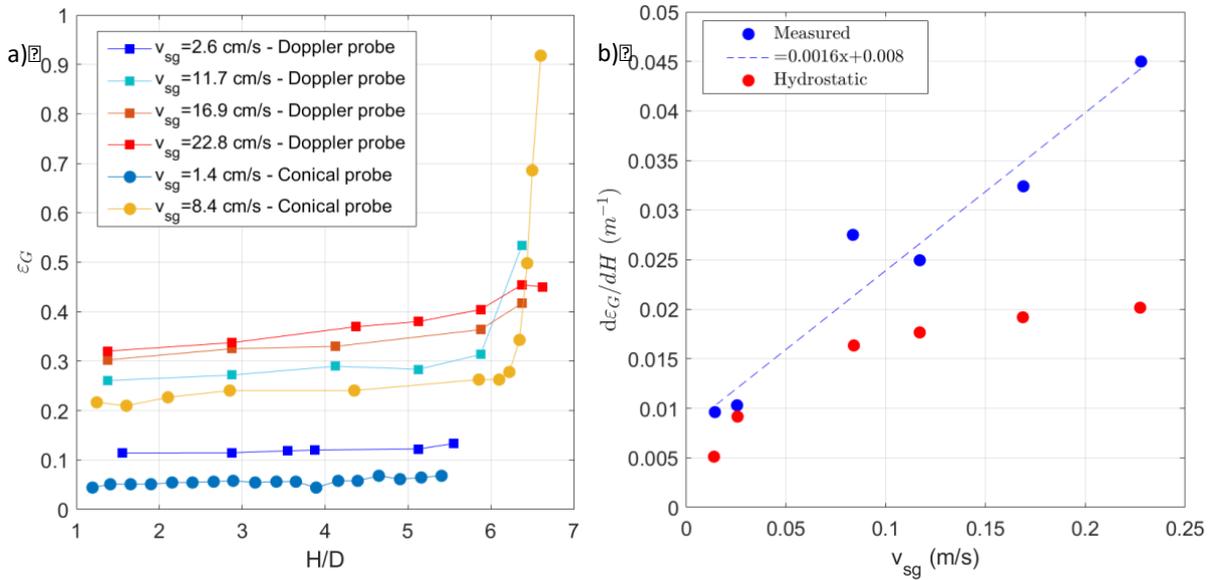

Fig.26: Evolution of the local void fraction measured on the column axis $\varepsilon_G(0)$ with the height H above injection (a), and evolution of the mean slope $d\varepsilon_G(0)/dH$ evaluated between H/D=1 and 6 versus the gas superficial velocity compared with $d\varepsilon_G/dH_{hydrostatic}$ (b). Measurements performed with a downward oriented Doppler probe.

The evolution of the local void fraction $\varepsilon_G(0)$ measured on the column axis at H/D=3.625 with the superficial velocity Vsg is provided Fig.27-a. According to that figure, the homogeneous regime ends up at Vsg about 4cm/s in our conditions. Beside, in the heterogeneous regime (at Vsg=9cm/s), the transverse void fraction profiles $\varepsilon_G(x/R)$ happen to be self-similar when normalized with $\varepsilon_G(0)$. That self-similarity holds (at least) from 2.6D above the injector up to 4.87D (see Fig. 27-b). In addition, the normalized profiles compare well with the fit proposed by Forret et al., 2006 that writes:

$\varepsilon_G(x/R) / \varepsilon_G(0) = - 1.1927 [(x/R)^6-1] + 0.8187 [(x/R)^4-1] - 0.6260 [(x/R)^2-1]$  (2)

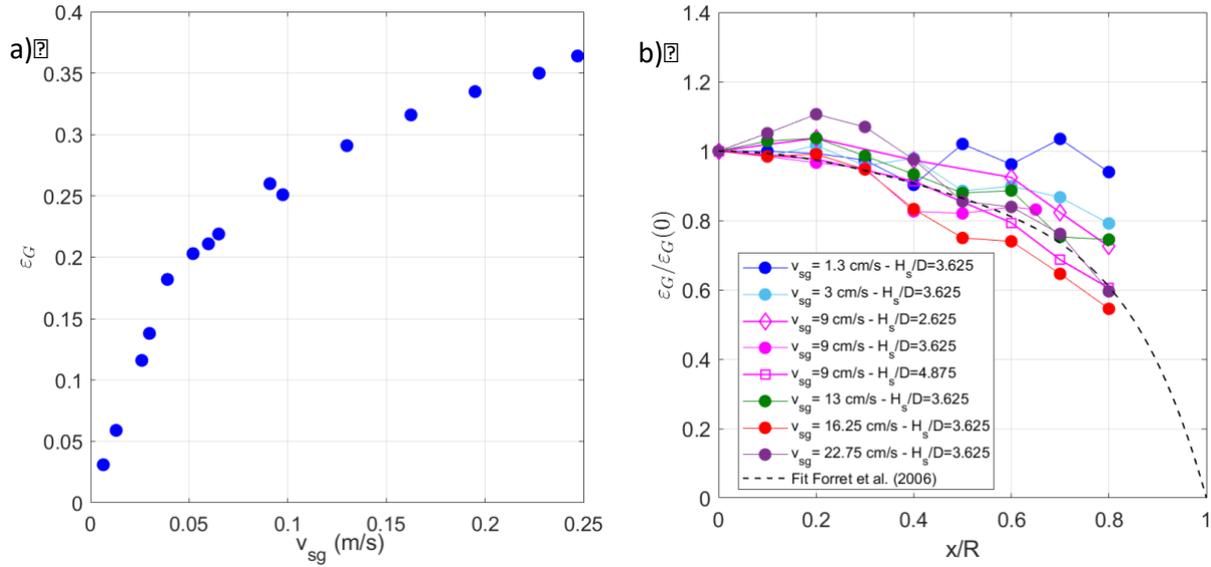

Fig.27: Local void fraction measured with a downward directed Doppler probe: a) evolution of the void fraction on the column axis with the superficial velocity at H/D=3.625 and b) transverse void fraction profiles normalized by the void fraction on the column axis measured at different heights above injection; comparison with the fit (eq.(2)) proposed by Forret et al., 2006.

4.3.2 *Bubble size*

Let us first examine the bubble chords distributions detected by a downward directed Doppler probe when considering direct velocity measurements only. As shown Fig.28-a, the chord pdf detected on the column axis and in the heterogeneous regime does not vary with the height above injection when in the quasi-fully developed region. That indicates that the size distribution was frozen in that region, that is either coalescence and break-up were not active, or their contributions equilibrate in such a way that the size distribution was left unchanged in the heterogeneous regime.

Then, by combining upward and downward orientations for the probe and using direct velocity measurements only, let us then examine how the mean chord $C_G$ and its standard deviation $C_G'$ evolve with the gas superficial velocity. According to Fig.28-b and 28-c, both quantities steadily increase in the homogeneous regime, and they both stabilize when in the heterogeneous regime. Indeed, the gas distributor has a fixed geometry so that the gas is ejected through the orifices with a velocity that increases with Vsg: the latter ranges from a few m/s up to ≈100m/s at Vsg=25cm/s. At the end of the homogeneous regime, the gas ejection velocity is about 20m/s, and the jet Reynolds number becomes of the order of ≈ 2000. This is roughly the onset of a turbulent jet. Hence, above the homogenous/heterogeneous transition, the bubble size distribution is mainly controlled by the break-up of the turbulent gas jets at injection, and that explains why it remains nearly the same in heterogeneous conditions. These bubbles have a mean chord comprised between 4 and 4.5mm (Fig.28-b) with a standard deviation that is about twice the mean (Fig.28-c).

In Fig.28-b and c, the mean chord and its standard deviation are also provided for a probe pointing downward (data series "up flow"), for a probe pointing upward (data series "down flow") and from the combined statistics (data series "up & down"). "Up flow" and "up & down" data almost coincide because of the very limited number of velocity data collected on the column axis with a probe pointing upward (the velocity validation rate in that case was at most 1.2%). Although limited, the data acquired with the probe pointing upward clearly indicate that the bubbles moving downward are much smaller in size: their mean chord is about 2 to 2.5mm to be compared with a mean chord about 4mm for bubbles moving up.

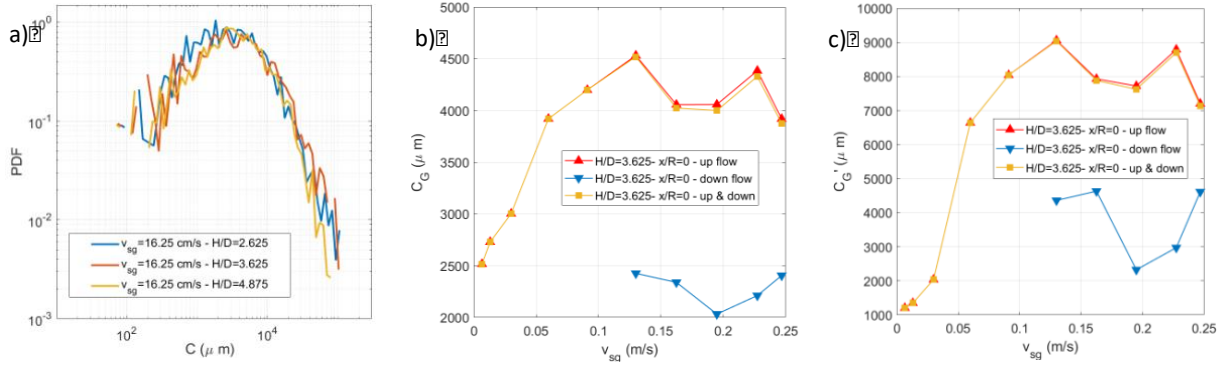

Fig.28: a) Chord distributions on the column axis at various heights above injection using a downward directed Doppler probe. Evolution with Vsg of the mean chord (b) and of the chord standard deviation (c) on the column axis and at H/D=3.625: comparison between statistics for up flow, down flow and combined flow directions (direct velocity measurements only).

Transverse profiles of the mean value $C_G$ and of the standard deviation $C_G$' of chords are provided Fig.29. The profiles of both $C_G$ and $C_G$' are nearly flat in the homogeneous regime (Fig.29-a). In the heterogeneous regime, a limited radial segregation appears with a slight decrease of the mean chord at large x/R: typically, for Vsg= 13cm/s and for Vsg=23cm/s; the chord decreases from ≈4mm in the center down to ≈3mm at x/R≈0.8 (Fig.29-b). In the mean time, the standard deviation decreases from ≈8mm in the center to 3-4mm at x/R≈0.8.

Fig.29-c compares mean values and standard deviations of chords between "up flow", "down flow" and "up & down" data series. These data confirm the difference in the mean chord between bubbles moving up and bubbles moving down. The standard deviation is also significantly larger (up to a ratio 2.5) for bubbles moving up compared with those moving down. These differences persist when changing Vsg while staying in the heterogeneous regime (not shown). Globally, the values from the "up flow" series are very similar to those from the "up and down flow" series: they are slightly higher because of the limited proportion of downward moving bubbles, a trend that holds up to x/R about 0.7-0.8.

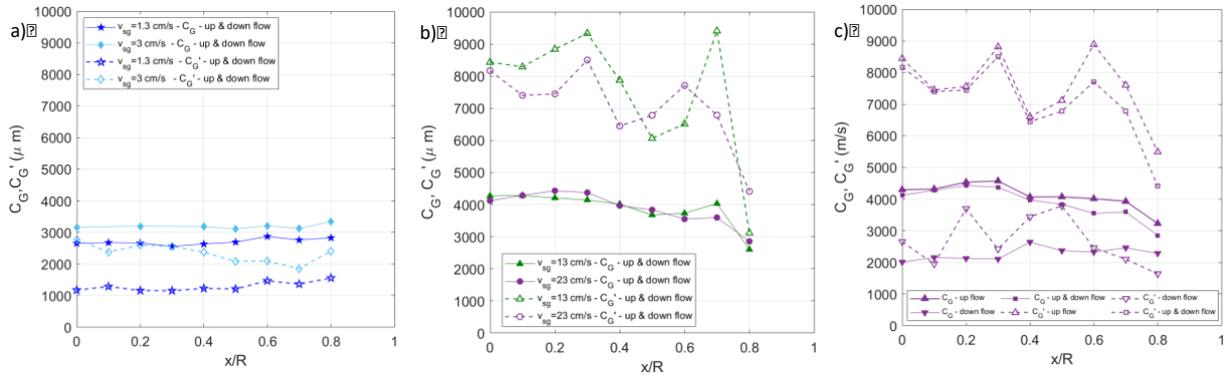

Fig.29: Transverse distributions of the mean chord and of its standard deviation measured with a Doppler probe: a) in the homogeneous regime at Vsg=1.3cm/s and 3cm/s from combined flow directions, b) in the heterogeneous regime at Vsg= 13cm/s and 23cm/s from combined flow directions, (c) comparison between moments for up flow, down flow and for combined flow directions at Vsg=23cm/s (direct velocity measurements only). Measurements at H/D=3.625.

Finally, let us compare the size measured with the Doppler probe with alternate techniques. Table 7 provides the Sauter mean vertical diameter measured with a single multimode optical probe using the dewetting time, as well as the Sauter mean horizontal diameter measured with the correlation technique (Maximiano Raimundo et al., 2016). From these data, we evaluate the mean eccentricity χ and the volume equivalent diameter Deq. From the arithmetic mean chord $C_G$ measured with the Doppler probe, a Sauter mean vertical diameter is deduced assuming spherical inclusions using $Dv_{32}=3C_G/2$. For ellipsoidal bubbles of the same eccentricity χ, that relation becomes $Dv_{32}=(3/2) C_G/\chi$ (Liu and Clark, 1995). We compare the $Dv_{32}$ estimated from the Doppler probe with Deq in the last two columns assuming either χ=1 or χ=0.85. If one except the data collected in the homogeneous regime at Vsg=1.3cm/s (in this flow condition, the perturbation induced by the probe holder is quite significant), the agreement between $Dv_{32}$ and Deq is satisfactory (15-20% deviation) when assuming spherical bubbles (i.e. χ=1). The agreement is even better (5% deviation) when considering a 0.85 eccentricity. In the present experiments, the mean equivalent bubble diameter remains the same throughout the heterogeneous regime: it equals 7mm±0.3mm. That size corresponds to distorted wobbling bubbles those terminal velocity $U_T$ in water and under earth gravity is about 0.22cm/s: their particulate Reynolds number is about 1500.

| vsg (m/s) | Mean chord from Doppler probe (mm) | Dv (mm) measured with a conical probe | Dh (mm) measured with correlation technique | Deq deduced from Dv and Dh (mm) | Mean eccentricity C = Dv/Dh | Dv32 estimated from Doppler probe with C=1 / Deq | Dv32 estimated from Doppler probe with C=0.85 / Deq |
|---|---|---|---|---|---|---|---|
| 0.013 | 2.51 | 6.66 | 6.6 | 6.62 | 1.01 | 0.58 | 0.69 |
| 0.065 | 3.65 | 6.69 | 6.82 | 6.77 | 0.98 | 0.81 | 0.95 |
| 0.13 | 3.97 | 6.47 | 7.35 | 7.05 | 0.88 | 0.84 | 0.99 |
| 0.163 | 4.0 | 6.22 | 7.59 | 7.1 | 0.82 | 0.85 | 0.99 |
| 0.195 | 4.0 | 6.59 | 7.76 | 7.35 | 0.85 | 0.82 | 0.97 |

Table 7: Comparison of mean bubble sizes measured with different techniques. Measurements taken on the axis at H/D=3.625. For the Doppler technique, the mean chord $C_G$ provided here was obtained from direct velocity measurements (no interpolation) from a downward directed probe.

4.3.3 *Mean bubble velocity*

Figure 30 provides the radial distributions of the mean vertical bubble velocity for two Vsg in the homogeneous regime and for three Vsg in the heterogeneous regime. The data series indicated "up flow" correspond to the mean vertical bubble velocity measured with a downward oriented probe only (and without velocity interpolation). The mean bubble velocity indicated as "up & down" corresponds to the neat vertical bubble velocity deduced by combining up and down probe orientations and by considering direct velocity measurements only (no interpolation). By convention, a positive velocity is directed against gravity, and a negative value is along gravity. The "up flow" and "up & down flow" velocity measurements are nearly the same in the homogeneous regime. In the heterogeneous regime, they coincide well in the center of the column where the flow is essentially upward directed, but they start to differ at x/R≥0.5: the difference between the two increases with the distance to the axis because of the presence of stronger downward directed motions as one approaches the wall as shown by the evolution of the bimodal pdfs in Fig.31.

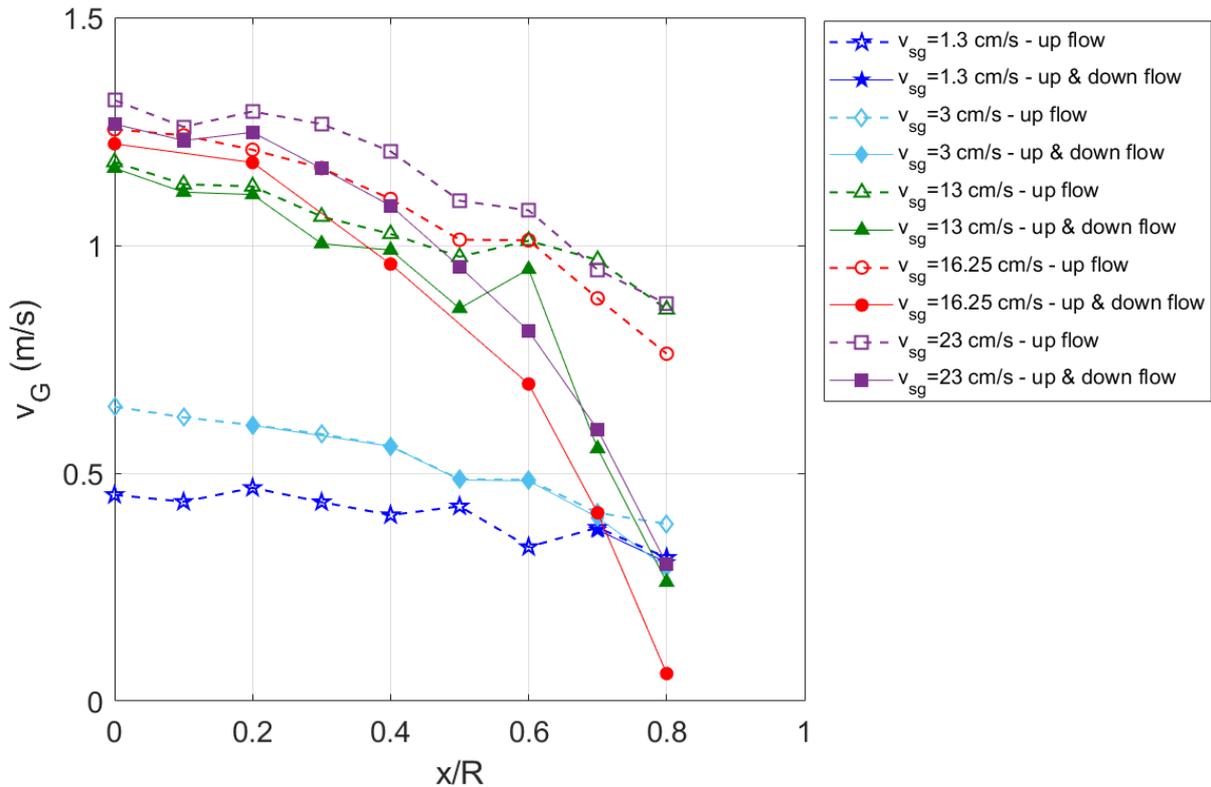

Fig.30: Radial profiles of the upward vertical mean bubble velocity for various Vsg as measured from *combined* (using up and down probe orientations) and *direct* (no interpolation) velocity measurements with the Doppler probe (indicated as "up & down flow" in the legend). Comparison with the mean vertical bubble velocity measured with a downward oriented probe only without interpolation (indicated as "up flow" in the legend). Measurements in the D=0.4m bubble column at H/D=3.625.

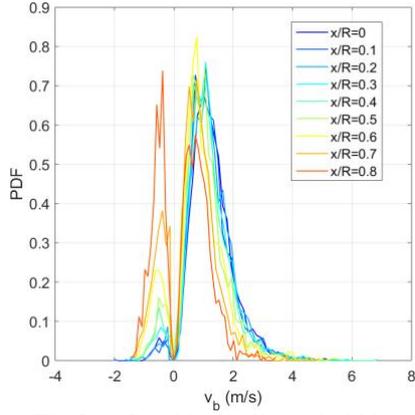

Fig.31: Combined direct bubble velocity pdfs at different lateral positions. Measurements in the D=0.4m bubble column at H/D=3.625 for Vsg=22.75 cm/s.

Globally, the bubble velocity profiles are rather flat in the homogeneous regime, and they become much steeper in the heterogeneous regime. We checked the self-similarity of bubble velocity profiles by normalizing them with the mean bubble velocity measured on the column axis. First, as shown Figure 32 for a fixed height (here H/D=3.625), the self-similarity is fulfilled for the combined direct mean bubble velocity at gas superficial velocities Vsg ranging from 8cm/s to 23cm/s, i.e. when in the heterogeneous regime. In addition, that the self-similarity also holds at different heights in the quasi fully developed region of the flow: that was tested for H/D between 2.63 and 4.62 at Vsg=9cm/s and at Vsg=16.3cm/s (not shown). For these flow conditions, we propose the following fit for the upward directed mean bubble velocity:

$$V_G(x/R) / V_G(0) = 1 - [(x/R) / 0.85]^3 \qquad \text{for x/R from 0 to 0.85} \qquad (3)$$

Eq.(3) works reasonably well, with an average deviation from experimental data about 14% and a maximum deviation of 27%. Let us underline that only the neat upward directed (positive) velocity is represented in Fig.32 and by the above eq.(3); additional measurements are needed between 0.85R and R to determine the downward directed mean bubble velocity that occurs in the near wall region. Contrary to the heterogeneous regime, the mean bubble velocity in the homogeneous regime remains upward directed over the entire cross-section (Fig.56). Let us also underline that, while the mean liquid velocity changes its direction at x/R ≈ 0.7-0.71 (see e.g. Forret et al., 2006; Maximiano Raimundo et al., 2019), Fig.32 indicates that the mean gas velocity crosses zero further away from the axis, at x/R≈0.85, than the mean liquid velocity for which x/R≈0.7. We retained the value 0.85 in eq.(3). Note that the relative locations of the mean flow inversion for the liquid and for the gas phases are consistent with the positive (i.e. upward directed) mean relative velocity of bubbles with respect to the liquid.

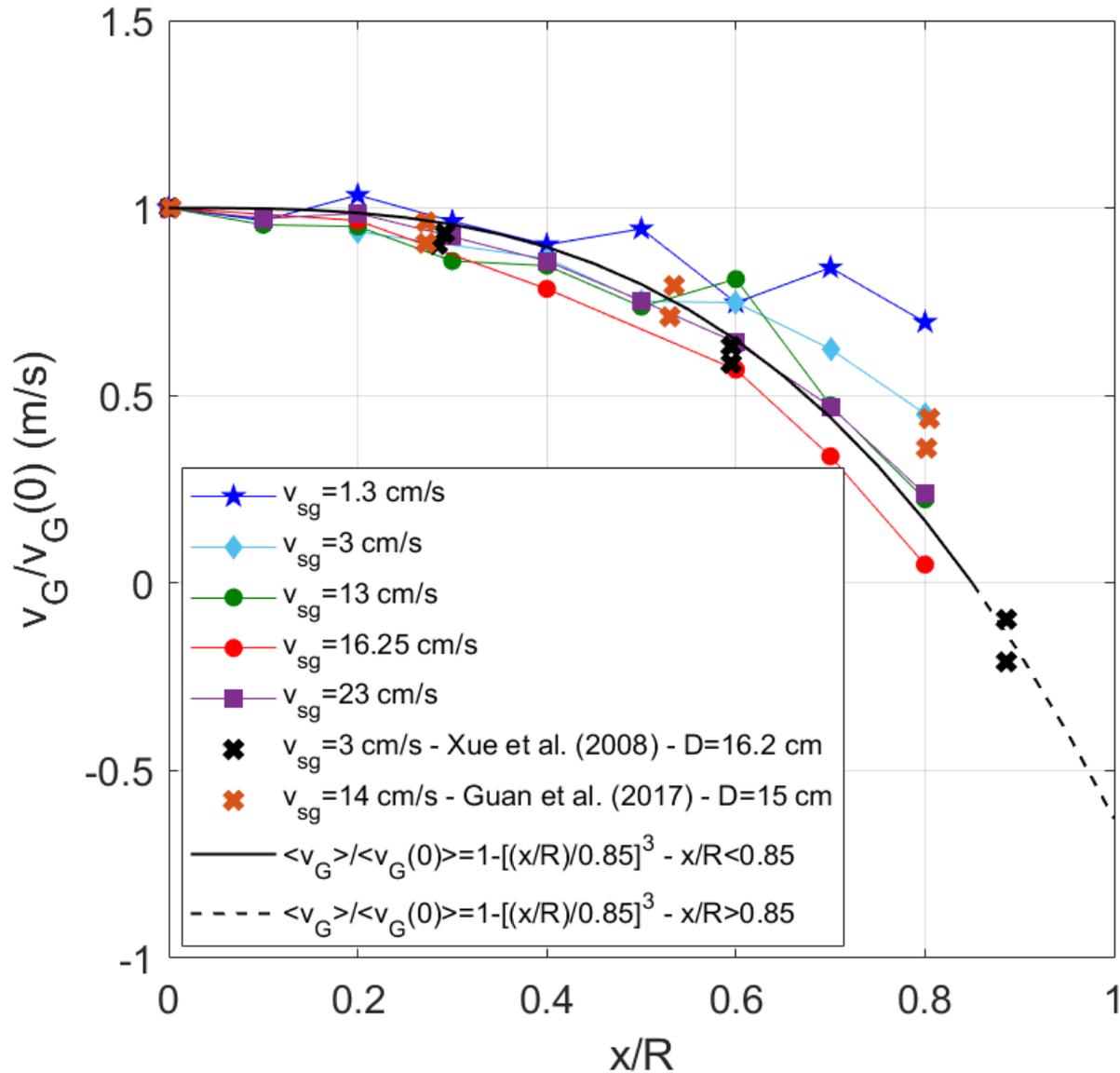

Fig.32: Radial profiles of the combined and direct mean bubble velocity scaled by the velocity on the axis measured in the D=0.4m bubble column at H/D=3.625. Comparison with eq.(3). Comparison with the data collected by Xue et al., 2008 with a four-tip probe at Vsg=30cm/s and H/D=8.5 in a D=0.162m diameter column. Comparison with the data collected by Guan et al., 2017 with a bi-probe at Vsg=14cm/s and H/D=5.3 in a D=0.15m diameter column.

It is interesting to compare the above findings with previous data, even though explorations of bubble velocity in the near wall region are scarce.
- Yao et al., 1991 performed measurements at H/D=12 in a D=0.29m column (deionized water and air, ambient conditions, no or weak coalescence, weak superficial liquid velocity of 1cm/s). With a five-point conductivity probe, they measured up to x/R=0.82 and, at Vsg=8cm/s (i.e. in the heterogeneous regime), they detected positive velocities only for the gas because they considered only one probe orientation. But, with the ultrasound technique that is sensitive to the flow direction, they indeed measured a downward bubble velocity at x/R=0.9 and for Vsg=8cm/s: this is in qualitative agreement with our results.
- Camarasa et al., 1999 measured gas velocity profiles with an ultrasonic probe in a D=0.1m column (tap water and air, ambient conditions, possibly some coalescence) for a porous plate injector that delivers bubble from 1 to 10mm in diameter. The measuring height above injection is not precisely indicated but it seems to be within the quasi-fully developed region. They observed negative gas velocities at x/R=0.75 in the heterogeneous regime (at Vsg=8.4cm/s and 10.7cm/s) on one side of the column, but positive velocities up to x/R=0.91 on the other side: clearly their mean flow was asymmetrical.

- Chen et al., 2003 in their 0.4m and 0.8 m diameter columns (tap water and air, ambient conditions, possibly some coalescence) exploited the Doppler signal collected with a cleaved optical probe. Their measurements were performed at H/D=2.6 for the 0.4 m column and at H/D=1.3 for the 0.8 m column: the latter condition is possibly not within the quasi-fully developed region. As they successively used upward and downward directed probes, negative bubble velocities were unambiguously detected. In the D=0.4 m column, the flow was axisymmetric, and negative velocities were observed between x/R≈0.6 up to x/R≈0.8 and the wall: these figures are comparable to our observations. A more quantitative comparison is not feasible, because the authors have not estimated the mean bubble velocity from the data collected with the two probe orientations. In the D=0.8 m column, Chen et al., 2003 observed downward motions between x/R≈0.4 up to x/R≈0.7 and the wall. However, as shown by the profiles of bubble detection frequency, the flows in that column were strongly asymmetrical (even at Vsg=3.3cm/s) and the authors attributed that to the presence of vortices at the column scale: whatever the reason, such conditions are not comparable to ours.
- Xue et al., 2008 performed bubble velocity measurements in a D=0.162 column (tap water and air, ambient conditions, probable significant coalescence) with a four-tip probe that was successively facing up and facing down. Although they do not clearly indicate how a mean velocity was deduced from these two sets, they consistently observed negative velocities (from ≈ -5 to -20cm/s) at x/R=0.9 on one side of the column when in the heterogeneous regime (from Vsg=8 to 60cm/s). They also consistently observed positive velocities (from ≈ 5 to 20cm/s) at x/R=0.9 on the other side of the column. Hence, and up to H/D≈5.1, all the heterogeneous conditions they considered (that is for Vsg between 8 and 60cm/s) correspond to strongly asymmetrical flows, a feature also confirmed by the void fraction profiles. At a larger distance from injection, (H/D=8.5) they measured one gas velocity profile that proves to be more symmetrical, with a neat positive velocity (≈ 50 cm/s) at x/R=0.6 and a neat negative velocity (between -10 and -20cm/s) at x/R≈0.88. These last observations are consistent with ours. In addition, the gas velocity profile they obtained at H/D=8.5 compares well with our data and with the fit we propose (see Figure 32).
- Guan and Yang, 2017 performed measurements in a D=0.15 column (tap water and air, ambient conditions, no or weak coalescence) with a horizontal dual conductivity probe. At H/D=5.3, they observed significant positive bubble velocities (between 0.3 to 0.5m/s) up to x/R=0.8 (Fig.32). Yet, they do not detail their signal processing, and it is unlikely that positive and negative velocities were distinguished by their measuring system. Thus, without additional detail on the signal processing, the results they provide on the gas velocity profiles in the near wall region should be taken with caution.

Overall, if one leaves aside the experiments in which asymmetrical flows were observed, literature results available on bubble velocity profiles are in qualitative agreement with our measurements.

*4.3.4 Standard deviation of the bubble velocity*

Transverse profiles of the standard deviation $V_G$' of the bubble vertical velocity are given Figure 33 at various gas superficial velocities. In the homogeneous regime (Vsg=1.3cm/s and 3cm/s), the profiles are flat over the entire cross-section. Note that a weak contribution due to bubbles moving downward does exist at large x/R. Such a contribution is probably due to some unsteadiness at injection because of gas maldistribution (Nedeltchev, 2020). An argument in favor of that scenario is that the downward contribution occurs only at very large x/R (≥0.7) when the gas flow rate is low (Vsg=1.3cm/s), while it is already present at x/R≈0.4 for a larger gas flow rate (Vsg=3cm/s).

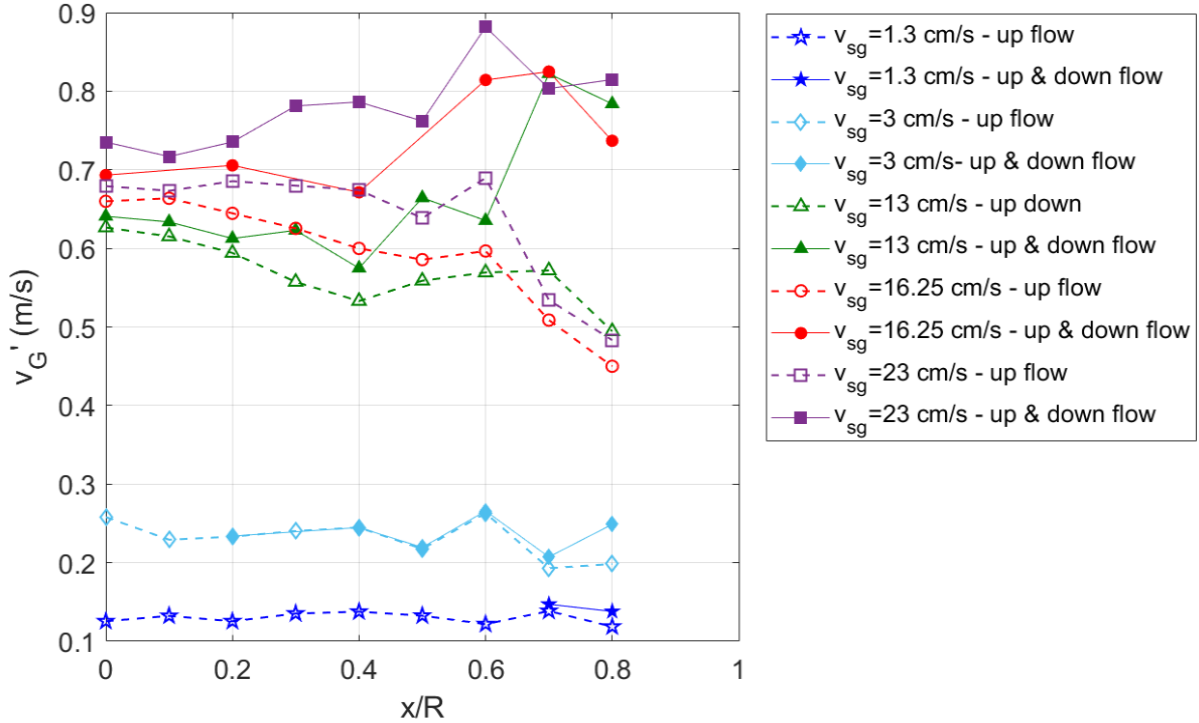

Fig.33: Transverse profiles of the standard deviation $V_G'$ of the bubble axial velocity estimated from combined and direct measurements (data series indicated as "up & down flow") and from direct measurements with a downward directed probe only (data series indicated as "up flow"). Measurements with the Doppler probe in the D=0.4m bubble column at H/D=3.625.

In the heterogeneous regime, the profiles of $V_G'$ are similar at all the gas superficial velocities considered, namely 13cm/s, 16cm/s and 22.75cm/s. The standard deviation $V_G'$ relative to the downward motion (data series "down flow" in Fig.33) is about 0.2 to 0.3 m/s: it is rather uniform. The standard deviation $V_G'$ relative to the upward motion (data series "up flow" in Fig.33) is larger (≈ 0.65 to 0.7m/s): it is maximum on the column center and it decreases radially down to 0.4-0.5m/s. Finally, the standard deviation $V_G'$ resulting from the combination of up and down motions (data series "up & down flows" in Fig.33) increases slightly from 0.7 to 0.8 m/s with the radial coordinate: that behavior is due to the increasing contribution of the downward oriented motion when approaching the column wall (see also the velocity pdfs in Fig.31).

The standard deviation $V_G'$ of the bubble vertical velocity on the column axis relative to the mean $V_G$ is plotted versus Vsg in Figure 34: the statistics presented here are based on combined and direct velocity measurements. For comparison, the quantity $V_L'/V_L$ evaluated on the column axis is plotted in the same figure. The mean $V_L$ and the standard deviation $V_L'$ of the liquid phase velocity were deduced from Pavlov tube measurements: these moments were evaluated considering the whole liquid velocity distribution (i.e. positive as well as negative velocities, see Fig.22). Both $V_G'/V_G$ and $V_L'/V_L$ vary with the gas superficial in the homogeneous regime. They both stabilize to nearly constant values in the heterogeneous regime: the velocity fluctuation relative to the mean reaches 0.55 in the gas, to be compared with ≈0.36 for the liquid.

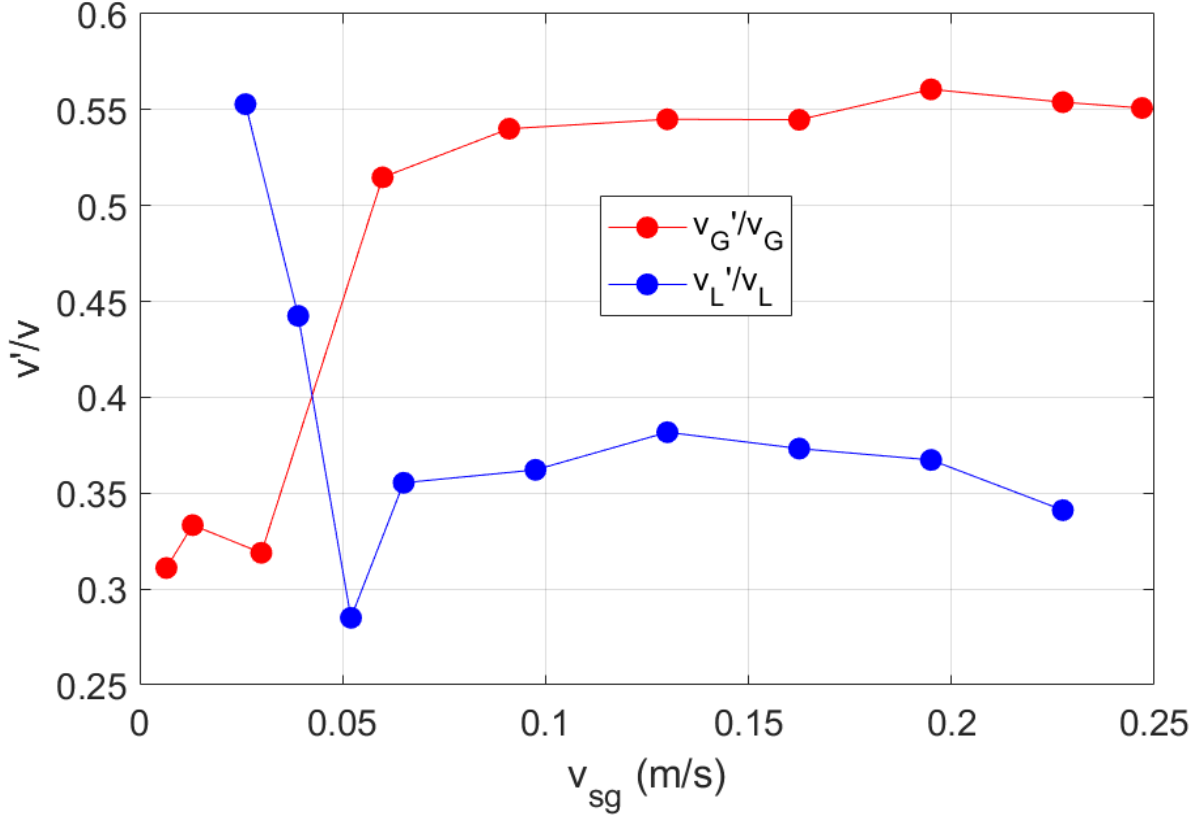

Fig.34: Evolution with Vsg of $V_G'/V_G$ measured on the axis of the D=0.4m bubble column at H/D=3.625 with a Doppler probe using combined and direct velocity measurements. Comparison with $V_L'/V_L$ measured at the same location with a Pavlov tube.

4.3.5 *Local and global gas flux, and comparison with the injected gas flow rate*

In that section, we discuss the global gas flux $Q_{Gup}$ flowing upward in the column as estimated from local measurements using various approaches that are detailed hereafter. All measured gas flow rates $Q_{Gup}$ are plotted versus the injected gas flow rate $Q_G$ in Figure 35. The gas flow rate $Q_G$ was measured with a battery of Brooks mass flow meters: the relative uncertainty less than 5% for all the flow conditions considered, and it was better than 2% in the heterogeneous regime.

A usual way to evaluate the local gas flux $j_G$ detected by local probes is to consider $j_G = \sum_i C_i/t_{tot}$, where the summation holds over all chords detected by the probe during the measuring duration. That formula corresponds to the expression of the gas flux in the classical two-fluid framework in which the local gas flux is the product of the local void fraction by the gas phase average velocity. In the heterogeneous regime, two difficulties arise with that approach. The first one is that the percentage of direct velocity (and thus of chords) detection is rather low as it limited to 50-60% of detected bubbles (see Section 3.3.1). The second, more drastic, difficulty is due to inclusions moving up and down. Indeed, with the present sensor, and as discussed in Section 4.1, it is not possible to identify the flow inversions from probe signal only. Consequently, we have no reliable interpolation scheme for the missing velocity data. Without surprise, when considering only direct velocity data to estimate the local flux, the global flux $Q_{Gup}$ resulting from the spatial integration of $j_G$ is underestimated (see closed squares in Fig.35): $Q_{Gup}$ represents only 30% of $Q_G$, meaning that the underestimation is huge in the heterogeneous regime.

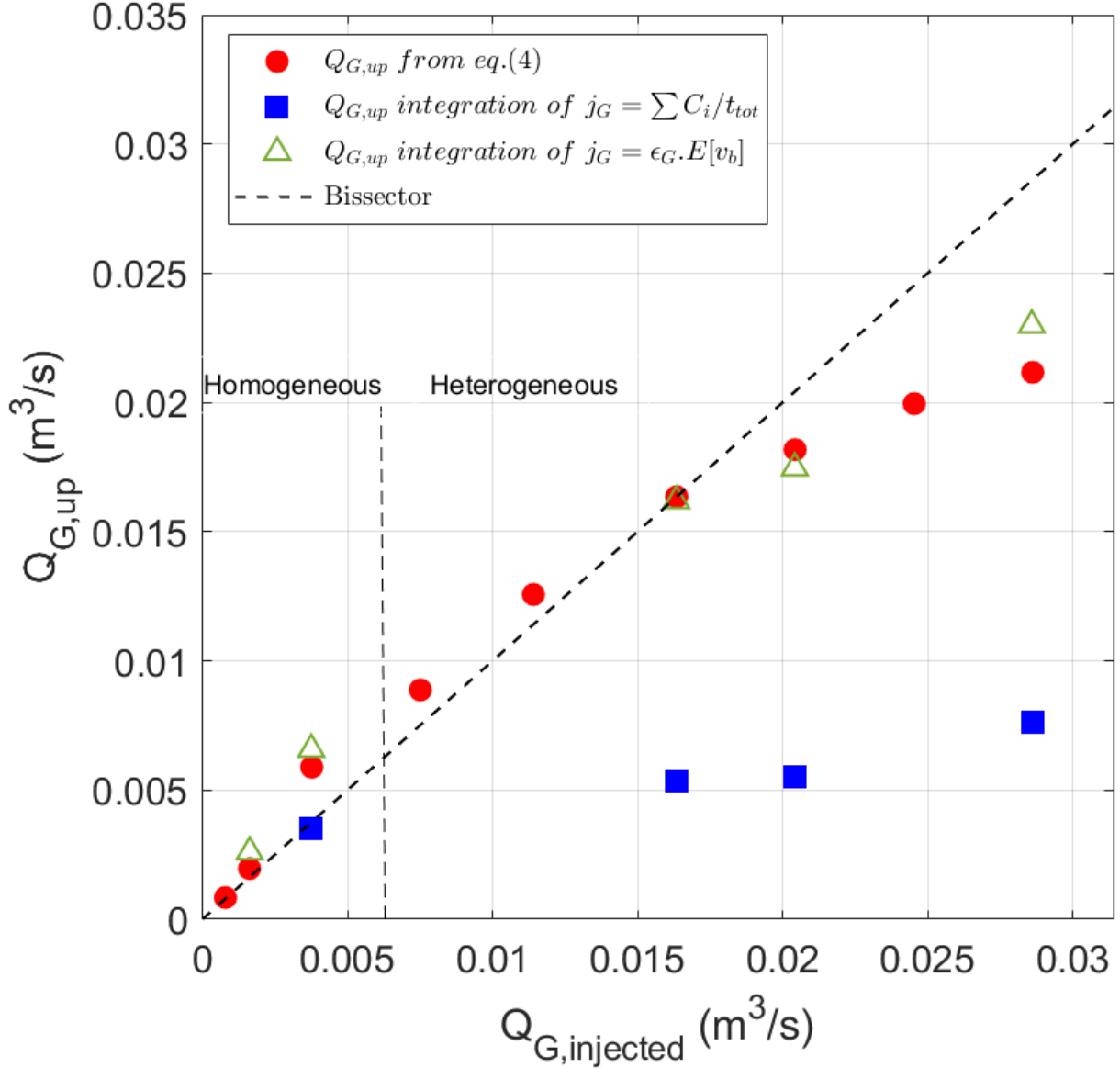

Fig.35: Upward gas flow rates deduced from the spatial integration of local fluxes measured at H/D=3.625 versus the injected gas flow rate. All data correspond to *combined* and *direct* velocity measurements. Local flux are evaluated either using $j_G = \sum_i C_i/t_{tot}$ (squares) or using $j_G = \varepsilon_G V_G$ (dots and triangles). Dots correspond to gas flow rates estimated using measurements on the column axis combined with eq.(4).

In order to circumvent the measuring difficulties related with changes in flow direction, we used a second approach based on an alternate way to evaluate the local gas flux. Indeed, in the framework of hybrid models inspired from kinetic theory, the dispersed phase is described by a product density $f^{(1)}(\mathbf{x}, D_b, V_b)$ where **x** refers to the position of the center of the gas inclusion, and where internal coordinates include the bubble size $D_b$, the bubble velocity $V_b$, plus possibly other bubble characteristics such as their eccentricity, etc… In that framework, the local gas flux is given by $j_G = \varepsilon_G E[V_b D_b^3]/E[D_b^3]$ where $E[.]$ is the ensemble average over the phase space (Cartellier, 1999). If bubble size and bubble velocity are not correlated, then $j_G = \varepsilon_G E[V_b]$ where $E[V_b]$ is the mean bubble velocity. Because of the very strong agitation of the liquid-gas mixture due to the presence of meso-scale structures (Maximiano Raimundo et al., 2019), bubble size and velocity are expected to be uncorrelated in bubble columns especially when in the heterogeneous regime provided that coalescence remains weak or absent. Besides, the mean bubble velocity $E[V_b]$ is readily available from the Doppler probe as the arithmetic average $V_G$ of bubble velocity measurements. The local void fraction required in the above formula is also available from the Doppler probe. Therefore, the local gas flux can be evaluated as $j_G = \varepsilon_G V_G$. The transverse profiles of the local flux $j_G = \varepsilon_G V_G$ evaluated from the mean bubble velocity (Figure 32) and from the void fraction (Figure 27) are shown Figure 36. As before, we focus on the upward flux, and we assume axisymmetric distributions for the spatial integration. Data are lacking for x/R>0.8, but, according to mean bubble velocity profiles (see Fig.32), the integration is stopped at x/R=0.85. The missing fraction of the upward flux for x/R between 0.8 and 0.85 has been estimated using plausible assumptions: as it represents less than a few percent of the global flux, it has been neglected. The global upward gas flow rates deduced from the integration of the transverse profiles are represented by triangles in Figure 35.

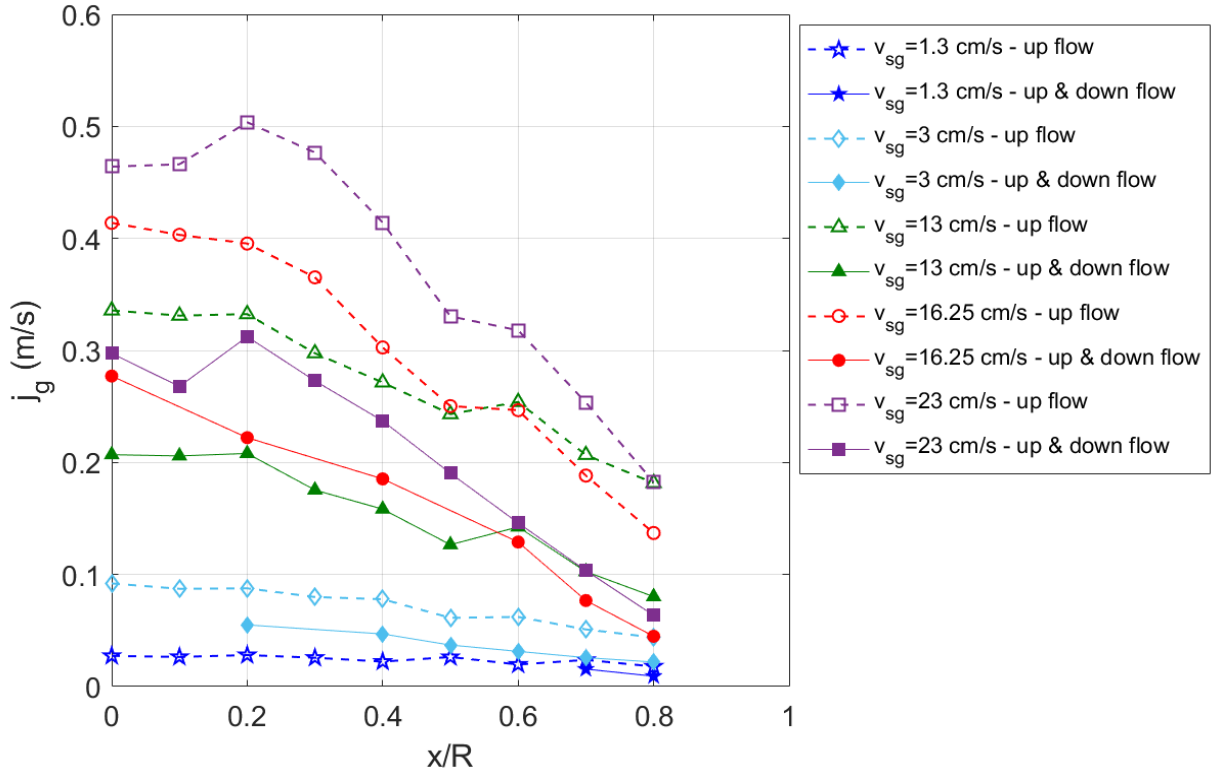

Fig.36: Local flux profiles evaluated with $j_g = \varepsilon_G V_G$, considering for the mean bubble velocity either *combined* and *direct* measurements (data series indicated as "up & down flow") or direct measurements with a downward directed probe only (data series indicated as "up flow"). Measurements at H/D=3.625 in a D=0.4m bubble column.

As a third approach, we considered the fits proposed in eq.(2) for the void fraction and in eq.(3) for the positive mean axial bubble velocity valid for the heterogeneous regime. The integration over the cross section of the product $\varepsilon_G V_G$ using these fits provides a relationship between the global gas flow rate $Q_{Gup}$ flowing upward in the column, and the values of the void fraction $\varepsilon_G(0)$ and of the mean, upward directed bubble velocity $V_G(0)$ on the column axis in the heterogeneous regime. For that regime, we obtain:

$$Q_{Gup} / (\pi R^2) = 0.3735 \, \varepsilon_G(0) \, V_G(0) \qquad (4)$$

The global gas flow rates evaluated from eq.(4) and from the measured values of $\varepsilon_G(0)$ (Figure 27) and of $V_G(0)$ (not shown) are represented in Figure 35 by closed dots.

Let us comment the results presented Figure 35. In the homogeneous regime (Vsg ≤ 5cm/s), the measured gas flow rate $Q_{Gup}$ should equal the injected gas flow rate $Q_G$. According to Figure 35, the measured gas flow rate is overestimated by at most 33%. A larger relative difference is observed at Vsg=3cm/s, but a maldistribution at injection occurs at this superficial gas velocity (see Section 4.3.4), and that condition cannot be considered as representative. To evaluate the sensor performance in terms of gas flux, more data with a better control of injection need to be gathered in the homogeneous regime.

In the heterogeneous regime, the global recirculation of the liquid entrains a fraction of the gas from the top to the bottom of the column. Therefore, the actual upward directed gas flow rate is expected to be larger than the injected gas flow rate $Q_G$. In addition, and owing to the observed increase of the void fraction with the height above injection (see Fig.26), the upward gas flow rate is also expected to increase with that distance. For these reasons, there is no undisputable reference available for the upward gas flow rate. We attempted to estimate the downward directed gas flow rate based on the experimental finding that the gas flux is directed downward for x/R between 0.85 and 1 (see Fig.32) and using different approximations for the transverse gas velocity profile.

- First, we tentatively assumed that the eq.(3) proposed above remains valid up to the wall. This is not demonstrated, but a couple of data from Xue et al., 2008 support that proposal (see Fig.32). Using that extended eq.(3) and the void fraction provided by eq.(2), the integral of the local flux $j_G = \varepsilon_G V_G$ within the downward directed zone represents 4.5% of the upward gas flow rate obtained by integrating between x/R=0 and 0.85.

- Second, assuming that the local bubble velocity $V_G$ equals $V_L$ plus a (constant) relative velocity $U_R$, and using the liquid velocity profile proposed by Forret et al., 2006, we compare the gas flow rate integrated between

x/R=0.85 and 1 to the gas flow rate integrated from the axis up to x/R=0.85. With $U_R$ set as the terminal velocity (which is a fixed quantity, independent from Vsg in our flow conditions – see section 4.3.2), the recirculated gas flow rate happens to represent between 1 and 4% of the upward gas flow rate. If one accounts for a swarm factor those effect is to increase $U_R$, these percentages would be even lower.

- Finally, if one us considers the crude limit in which $U_R$ is set to zero (i.e. no slip velocity), the recirculated gas flow rate fraction reaches at most 20%.

To appreciate the validity of the above estimates, we tested whether the fits proposed by Forret et al., 2006 for the liquid velocity and for the void fraction ensure equal upward and downward liquid fluxes. We found differences ranging from 20% (at transition) to -20% (at large gas hold-up): these figures are typical of the reliability of the proposed fits. Owing the various approximations considered above, it is difficult to unambiguously conclude on the magnitude of the recirculated gas flow rate. Tentatively, the later seems comprised between a few percent and 20% of the upward gas flow rate. Knowing these magnitudes, and despite the associated uncertainties, let us now comment on the difference between the measured upward global gas flux and the injected gas flow rate in the heterogeneous regime. According to Figure 36, at moderate Vsg, the measured gas flow rate exceeds $Q_G$ by at most 18%. Above Vsg=13cm/s, the trend becomes opposite. Moreover, the difference monotonously increases with Vsg, and it reaches -27% at the largest Vsg considered here (=24.7cm/s). Taking into account the unknown contribution of the recirculation, these differences with the actual upward gas flow rate are probably larger. At this stage, it is difficult to unambiguously identify the origin of the differences between $Q_G$ and $Q_{Gup}$. On one hand, the perturbation induced by the thick probe holder used to limit vibrations is possibly significant in heterogeneous conditions. On another hand, the characteristics of the global recirculation inside a bubble column are not known with enough accuracy. In that perspective, it would be worth to closely scrutinize the near wall region using a Doppler probe inserted in the flow through the wall to avoid vibration issues and to minimize the flow perturbation occurring in these very low, and thus sensitive, velocity zones. At the present stage of the investigation, we can nevertheless recommend to evaluate the upward directed gas flux based on $j_g = \varepsilon_G V_G$, with the void fraction measured with a downward oriented probe combined with a mean bubble velocity deduced from the combination of direct velocity measurement for a probe pointing downward and then pointing upward. With such an approach, one gets rid of the issue on data interpolation that is not solved for a flow that fluctuates between up and down motions.

## 5. Conclusions and perspectives

A new optical probe for size and velocity measurements of gas inclusions has been designed based on a single mode optical fiber with an ad-hoc tip geometry. That probe combines the detection of incoming interfaces at a high spatial resolution (its latency length is 6µm) with absolute velocity measurements from the analysis of Doppler bursts occurring before gas-liquid transitions.

From the analysis of its optical response and from experiments in controlled conditions, we have shown that the Doppler probe is sensitive to the trajectory of the incoming inclusion. More specifically, it only detects an interface approaching the probe tip head-on, and those normal makes an angle with the fiber axis less than about 12-13°. That angular limit can be slightly changed by varying the number of oscillations and/or the threshold in amplitude considered. Thanks to such sensitivity, the Doppler probe provides the translation velocity of the gas inclusion projected along the fiber axis. The uncertainty on velocity measurement is less than 14%.

An optimized signal processing has been developed to extract variables of interest, namely gas arrival times, gas residence times and velocities, from which one can infer statistics on velocity and size as well as concentration, volumetric flux, number density, number density flux… The sensor and the processing were tested in an air-water bubble column of diameter 0.4m and for superficial gas velocities between 1.3cm/s and 25cm/s. No bias related with bubble size or with bubble acceleration was identified. The percentage of bubbles detected for which a direct velocity measurement is available happens to be 66% in the homogeneous regime. That success rate drops down to 50% in the strongly unsteady and three-dimensional flows encountered in the heterogeneous regime. Yet, in these flow conditions, the mean bubble size measured with the Doppler probe was in agreement with alternate techniques.

The sensor was exploited to gather new information on bubble motion in the heterogeneous regime. Over this regime, the mean equivalent diameter remained constant, equal to 7mm±0.3mm, so that the corresponding particle Reynolds number based on the terminal velocity was also fixed, equal to about 1500. As for the liquid, the transverse profiles of the mean bubble velocity were found to be self-similar in the quasi-fully developed region of the column. A fit is proposed for the upward directed mean bubble velocity: the latter reaches zero at a lateral position x/R≈0.85. The transverse profiles of the standard deviation of the bubble velocity are nearly flat. Meanwhile, on the column axis, the standard deviation of bubble velocity with respect to the mean increases with the gas superficial velocity in the homogeneous regime, and it remains constant, equal to 0.55, in the entire heterogeneous regime. This is significantly larger than the ratio 0.36 found for the liquid phase.

In terms of perspectives, future efforts should be devoted to a refined investigation of near wall zones to quantify the recirculated gas flow rate as a function of flow parameters. To avoid the perturbation induced by the large probe holder used in the present experiments, new data should be preferentially collected using a Doppler probe inserted from the wall. Also, better resolved liquid velocity measurements are worth to be undertaken using less invasive techniques. Finally, to better understand the hydrodynamics of bubble columns, the Doppler probe opens the way to reliable bubble velocity measurements conditioned by the local concentration. Indeed, as the existence of clusters and of void regions has been evidenced from a Voronoï analysis of the phase indicator function delivered by optical probes (Maximiano Raimundo et al., 2019), information conditioned by the void fraction are crucial to understand the hydrodynamics of bubble columns. One objective in particular is to relate the swarm factor with the internal flow organization for given column size and injected gas flow rate.

**CRediT authorship contribution statement**
Anthony Lefebvre: Investigation, Methodology, Data curation, Writing - original draft
Yann Mezui: Investigation; Methodology, Data curation, Visualization
Martin Obligado: Investigation, Software, Data curation
Stéphane Gluck: Conceptualization , Investigation, Supervision
Alain Cartellier: Conceptualization, Investigation, Methodology, Roles/Writing - original draft; Writing - review & editing, Supervision.

**Declaration of Competing Interest**
The authors declare that they have no known competing financial interests or personal relationships that could have appeared to influence the work reported in this paper.

**Acknowledgments**
The LEGI is part of the LabEx Tec21 (Investissements d'Avenir  - grant agreement n° ANR-11-LABX-0030). That research was also partially funded by IDEX UGA (n° ANR-15-IDEX-0002).

**Nomenclature**

*Abbreviations*
| | |
|---|---|
| ADC | analog to digital converter |
| I.D. | internal diameter |
| LDV | Laser Doppler Velocimetry |
| pdf | probability density function |
| RPP | random Poisson process |
| 0C | cleaved tip |
| 1C | conical tip |
| 3C | conical-cylindrical-conical tip |
| 1D | one dimensional |
| 3D | three dimensional |

*Latin letters*
| | |
|---|---|
| a | bubble radius, m |
| C | chord cut through a bubble, m |
| $C_i$ | chord of the i$^{th}$ bubble, m |
| $C_G$ | mean chord, m |
| $C_G'$ | standard deviation of chord pdf, m |
| $C_{AM}$ | added mass coefficient, - |
| d | fiber core diameter, m |
| $d_c$ | probe tip size, m |
| $d_{probe}$ | probe diameter, m |
| D | column diameter, m |
| $D_b$ | bubble diameter, m |
| $DR_{measured}$ | measured dynamic range, - |
| $DR_{filter}$ | dynamic range set by filters, - |
| $DR_0$ | actual dynamic range, - |
| f | frequency, Hz |
| $f_D$ | Doppler frequency, Hz |
| $f_k$ | instantaneous frequency, Hz |
| $f_{Dmin}$ | minimum Doppler frequency, Hz |
| $f_{Dmax}$ | maximum Doppler frequency, Hz |
| $f_{HP}$ | high pass frequency, Hz |
| $f_{LP}$ | low pass frequency, Hz |
| $f_{sampling}$ | sampling frequency, Hz |
| $f_0$ | light frequency in vacuum, Hz |
| $f_B$ | bubble detection frequency, Hz |
| g | gravitation acceleration, m/s$^2$ |
| $H_0$ | static liquid height, m |
| H | height above injection, m |
| I, $I_1$, $I_2$ | light intensities, W/m$^2$ |
| $I_0$ | light intensity at waist, W/m$^2$ |
| $j_G$ | local gas flux, m/s |
| **k** | unit vector along the light direction of propagation, - |
| L | length, m |
| $L_s$ | latency length, m |
| $L_G$ | cumulated gas length, m |
| $n_1$, $n_2$, $n_{ext}$ | refractive indexes, - |
| $n_{eff}$ | effective group index, - |
| n | number of Doppler periods, - |
| **n** | vector normal to the interface, - |
| ☐☐ | numerical aperture, - |
| Q | flow rate, m$^3$/s |
| r | distance to the interface, m |
| $R_{a/b}$ | Fresnel coefficient at the frontier between media a and b, - |
| R | column radius, m |
| t | time, s |
| $t_A$, $t_E$ | characteristics times, s |
| $T_G$, $T_{Gi}$ | gas residence time, gas residence time of the i$^{th}$ bubble, s |
| $T_i$ | date of the center of the i$^{th}$ bubble, s |
| $t_{tot}$ | total measuring duration, s |
| $T_m$ | de-wetting time, s |
| $u_b$ | bubble velocity with respect to the probe, m/s |
| $U_T$ | terminal velocity, m/s |
| V | interface or reflector velocity, m/s |
| $V_{axis}$ | interface displacement velocity projected along the probe axis, m/s |
| $V_b$, $V_{bi}$ | bubble velocity, velocity of the i$^{th}$ bubble, m/s |
| $V_G$ | mean bubble velocity, m/s |
| $V_G$' | standard deviation of bubble velocity pdf, m/s |
| $V_{max}$ | maximum velocity detected, m/s |
| $V_{min}$ | minimum velocity detected, m/s |
| Vsg | gas superficial velocity, m/s |
| x | radial coordinate, m |
| $\Delta t_i$ | waiting time between successive direct bubble velocity measurements, s |
| $<\Delta t_i>$ | mean waiting time, s |
| $\Delta t_{cut}$ | threshold on the mean waiting time, s |
| $\Delta t_k$ | Doppler period, s |

*Greek letters*

| | |
|---|---|
| α | inclination angle, rad |
| β | collection angle, rad |
| χ | aspect ratio, - |
| ε | relative difference between periods, % |
| $\varepsilon_G$ | local void fraction, - |
| $\lambda_0$ | wavelength in vacuum, m |
| λ | wavelength in medium, m |
| ν | kinematic viscosity, m$^2$/s |
| θ | angle, rad |
| Θ | illumination angle, rad |
| $\Theta_{eff}$ | effective illumination angle, rad |
| ρ | density, kg/m$^3$ |
| σ | surface tension, N/m |

*Dimensionless numbers*

| | |
|---|---|
| M | modified Weber number, $[\rho_L D_b u_b^2/\sigma] (D_b/d_{probe})$ |

Re$_p$ particulate Reynolds number, $U_T D_b / \nu_L$

*Subscripts*
D  Doppler
G  Gas phase
L  liquid phase

_______________________